\documentclass[12pt,fleqn]{article}

\input{epsf}
\def\be{\begin{eqnarray}}
\def\ee{\end{eqnarray}}

\def\bra{\langle}    
\def\ket{\rangle} 

\def\Im{{\rm Im}\,}
\def\Re{{\rm Re}\,}

    \def\calV{{\cal V}}

\begin{document}
\begin{center}
{\Large{\bf CHIRAL UNITARY APPROACH TO THE}}
\end{center}
\begin{center}
{\Large{\bf  $\pi N^\ast N^\ast$, 
$\eta N^\ast N^\ast$ COUPLINGS FOR THE}}
\end{center}
\begin{center}
{\Large{\bf $N^\ast(1535)$ RESONANCE}}
\end{center}

\vspace{1cm}

\begin{center}
{\large{ J.C. Nacher$^ {1,2}$, A. Parre\~no$^ {3}$, E. Oset$^ {1,2},$}}
\end{center}
\begin{center}
{\large{A. Ramos$^4$, A. Hosaka$^{5}$ and M. Oka$^{6}$}}
\end{center}

\vspace{0.4cm}
%{\it $^1$  Departamento de F\'{\i}sica Te\'orica and IFIC 

%Centro Mixto Universidad de Valencia-CSIC
%46100 Burjassot (Valencia), Spain.}       

\begin{center}
{\it $^1$ Research Center for Nuclear Physics (RCNP), Osaka University,}
%\end{center}
%\begin{center}
{\it Ibaraki, Osaka 567-0047, Japan.}
\end{center}

\begin{center}
{\it $^2$  Departamento de F\'{\i}sica Te\'orica and IFIC, 

Centro Mixto Universidad de Valencia-CSIC
46100 Burjassot (Valencia), Spain.}
\end{center}

\begin{center}
{\it $^3$ Institute for Nuclear Theory, 
University of Washington, Seattle, U.S.A.}
\end{center}

\begin{center}
{\it $^4$ Departament d'Estructura i 
Constituents de la Mat\'eria, Universitat de

Barcelona,}
%\end{center}
%\begin{center}

{\it Diagonal 647, 08028 Barcelona, Spain.}

\end{center}

\begin{center}
{\it $^5$ Numazu College of Technology, 3600 Numazu, 410-8501, Japan}
\end{center}

\begin{center}

{\it $^6$ Department of Physics, Tokyo Institute of Technology, }

{\it Meguro, Tokyo 152-8551, Japan}
\end{center}
\vspace{2.2cm}

\begin{abstract}
{\small{Using a chiral unitary model in which the negative parity 
nucleon 
resonance $N^\ast \equiv N^\ast(1535)$ is generated
dynamically by means of the Bethe Salpeter equation with 
coupled meson baryon
channels in the $S = 0$ sector, we have obtained 
the $\pi^0 N^\ast N^\ast$ and $\eta N^\ast N^\ast$ couplings. 
The $\pi^0 N^\ast N^\ast$ coupling has
smaller strength but the same sign as the $\pi^0 N N$ coupling. 
This rules out
the mirror assignment of chiral symmetry 
where the ground state nucleon $N$ and the negative parity resonance 
$N^\ast$ are envisaged as chiral
partners in the baryon sector. 
%together with the $\pi$ and $\sigma$ in the mesonsector.
}}
 
\end{abstract}
\vspace{2.2cm}
\newpage

%==================================================================
\section{Introduction}
%==================================================================

Chiral symmetry has been playing 
a crucial role in understanding hadron physics.  
The current algebra and associated 
low energy theorems have been successfully applied to 
hadronic phenomena which in particular involve pions that appear 
as the Nambu-Goldstone bosons of spontaneous breakdown of chiral 
symmetry~\cite{coleman}.  
Recent developments of chiral perturbation methods for mesons and 
baryons are partly motivated by the hope in the theoretical side 
to describe hadrons without referring to particular 
models~\cite{1,2,3,4,5}.  

% Development of experimental facscilities at such as TJNAF, COSY and 
% SPring8 is also an important factor to push
% forward the physics along this line.   
% In aproaches based on the chiral perturbation theory 
% the main ingredients are mesons and 
% baryons in the ground state 
%which are stable against strong decays, 
%and chiral symmetry.    
In Refs.~\cite{6,7,8}, it was demonstrated that chiral perturbation 
theories, when unitarization in coupled channels is 
incorporated, can be applied up to resonance energy regions which 
are much beyond what the original chiral perturbation theories are 
supposed to be applied to.  
In Refs.~\cite{6,7} this approach was referred to as the chiral 
unitary approach.  
% In these methods, some observed excited states appear as 
% resonances of mesons and/or baryons in ground states.  

The advantage of the use of the chiral unitary approach is based on the 
implementation of exact unitarity 
together with a chiral expansion of Re($T^{-1}$) instead of the $T$ matrix, 
in a way analogous to the effective range expansion in quantum 
mechanics.  
These manipulations allow one 
to extend the information contained in the chiral lagrangians with 
small number of derivatives to higher energies than expected.

In this way all meson resonances in meson meson scatterings up to 1.2 GeV
were obtained \cite{6,7}. Simultaneously an alternative but equivalent unitary
approach was followed in Ref.~\cite{D} where the lowest order chiral Lagrangian
was explicitly kept and the higher order one was generated from  the
exchange of genuine QCD resonances, following the idea of Ref.~\cite{Pichmas}. This
allowed one to distinguish between genuine QCD resonances and other 
resonances which
come as a consequence of the scattering of the mesons.

While a great success has been achieved in explaining hadron 
phenomena, several fundamental questions 
are not yet clearly answered.  
Here we would like to raise the following particular one 
concerning the formation of chiral multiplets:
Are there any particles which belong to the same 
chiral multiplets and so they will get degenerate when chiral symmetry is 
restored?  
In a linear sigma model, for instance, of $SU(2) \times SU(2)$ chiral 
symmetry, a would-be chiral multiplet includes the $\sigma$ and 
$\pi$ mesons, which are the vector representations of 
the chiral group.  
In the spontaneously broken phase the pion becomes massless and the 
sigma meson remains massive, while in the symmetric phase they get
degenerate and acquire an equal mass. 
Another candidate of a chiral multiplet is the pair of the vector 
($\rho$)  and axial vector ($a_1$) mesons.  
In both cases, the mass difference between particles in the same 
multiplet is considered to be generated by a finite scalar condensate 
associated with spontaneous break down of chiral symmetry. 

For baryons, less attention has been paid in identifying them as 
members of chiral multiplets.  
For instance, one may ask 
what would be the chiral partner of the nucleon $N(939)$, 
and what properties are dictated by the underlying chiral symmetry.
In Refs.~\cite{joh1,joh2}, motivated by these questions, 
the chiral symmetry of 
the nucleon has been investigated, where 
two distinct chiral assignments for baryons were discussed.  
In one assignment, the positive and negative parity nucleons belong 
to different chiral multiplets, each of which is an 
independent representation of chiral symmetry.  
This case is referred to as the naive assignment. 
% , or the associated 
% nucleons as the naive nucleons (or representation).    
In the other case, positive and negative parity nucleons form 
a chiral multiplet, in which they transform to each other under chiral 
transformations.  
This case is referred to as the mirror assignment.
% , or the associated 
% nucleons as the mirror nucleons (or representation).  

In both assignments, chiral symmetry puts unique 
constraints on properties of the positive and negative parity 
nucleons.  
In the naive assignment, a chiral symmetric lagrangian reduces
to a sum of two lagrangians for positive and negative parity 
nucleons.   
Therefore, 
couplings between them disappear to leading order.  
In Ref.~\cite{joh2} this fact was considered to be the reason for
the small coupling constant $g_{\pi N N^*}$ for the decay of 
$N^* = N^*(1535)$.  
In contrast, in the mirror case, there are several 
interesting facts coming out.  
One of the non-trivial observations is that the positive and 
negative parity nucleons in the same chiral multiplet will get degenerate  
having a finite mass when chiral symmetry is restored.  
This and related properties lead to several interesting predictions 
on the behavior of the particle spectrum and interactions toward the 
restoration of chiral symmetry.  
For instance, the rate of meson production in nuclei depends 
crucially on the above chiral assignment~\cite{kjo}.  

One interesting signal which can be utilized to distinguish 
the two chiral assignments for the nucleon is 
the relative sign of the axial vector coupling constants.   
In the naive representation, their coupling constants take the same 
sign, while in the mirror representation they carry different signs.      
In reality, the physical nucleons can be combinations of 
the two representations, and therefore  
depending on the mixing rate their 
relative sign can be either positive or negative.  
If, however, the sign were be negative, the physical 
nucleons would contain to a large extent the mirror component.  

To our best knowledge up to date, 
the axial vector coupling constants of excited states
have not been studied
from the point of view of chiral symmetry.  
The main purpose of the present work is then to extract 
information theoretically 
on the sign of the axial vector coupling constant of the 
negative parity nucleon, $N^*(1535)$.  
We have chosen this state since it is the lowest negative parity state 
in nucleon excitations and would most likely be the chiral partner 
of the nucleon if any.  
Practically, it is convenient to compute the pion-baryon couplings, 
since they are related to the axial vector coupling constants 
through the Goldberger-Treiman relation.  
This is the main object in the present work.  
Since our interest is essentially related to chiral symmetry, it 
should be crucially important that the method adopted should respect 
chiral symmetry.  
In this respect, it is possible and looks appropriate to 
adopt a chiral unitary approach  
for the description of the negative parity nucleon $N^*(1535)$.

In fact, two chiral assignments can be distinguished in the linear 
representation of chiral symmetry.   
In the non-linear representation, the difference between 
them is masked by the unitary transformation 
$\psi \to B = u \psi$ (see eq. (2)), when both the positive and negative 
parity nucleons are introduced as elementary particles.  In the present 
work, however, the negative parity nucleon is generated dynamically 
through the elementary constituents of the positive parity (ground state) 
nucleon and mesons.  The nature of the negative parity nucleon is therefore
an interesting dynamical question.  Furthermore, 
the axial charge $\psi^\dagger \gamma_5 \psi$ is invariant under the 
transformation $\psi \to u \psi$, and therefore, this quantity is suitable 
to test the chiral symmetry of the nucleon.

A chiral unitary approach to the meson baryon problem has been done in
Refs. \cite{9,10,11,12} using the Lippmann Schwinger [LS] equation and input
from chiral Lagrangians. The $\Lambda(1405)$ and $N^*(1535)$ resonances
are generated dynamically in those schemes in the $S=-1$,  $S=0$ channels,
respectively. The connection between the inverse amplitude method
and the Lippmann Schwinger equation (or better, the Bethe Salpeter equation)
is done in Ref. \cite{6} , where a justification
is found why in some channels, like in the case of $\bar{K} N$ scattering
in $s$-wave, the LS equation using the lowest order chiral Lagrangian
and a suitable cut off in the loops can be a good approximation \cite{12}.
The procedure is not universal and in the case of $\pi N$ scattering and coupled 
channels explicit use of the higher order Lagrangians
is needed \cite{11}. Alternatively one can use dispersion relations with
different subtraction constants in each channel, which is the approach
followed here. In both cases 
the $N^\ast(1535)$ resonance is generated dynamically 
such that acceptable results for low energy cross sections 
for $\pi N\rightarrow \eta N$ and related channels can be obtained.

This paper is organized as follows.  
In section 2, we briefly discuss the coupled channels chiral 
unitary approach for the $N^*(1535)$ resonance.  
In section 3, we demonstrate how the strong couplings of the pion
and eta to the $N^*(1535)$ are derived.  
There appear many diagrams which contribute to the relevant 
couplings.  
We investigate properties of all of these diagrams in detail.  
Numerical results are presented in section 4.  
We will find there that the $\pi^0 N^\ast N^\ast$ coupling has
smaller strength and the same sign as the $\pi^0 N N$ coupling. 
In section 5 we compare the present results with other model 
calculations.  
The final section is devoted to a summary and concluding remarks.  

%  In the scalar sector, the importance of the unitarization of the lowest order
%  of $\chi PT$ amplitude, and the appearance of genuine QCD resonances at higher
%  energies, makes it possible to obtain a good reproduction of the data up to
%  1.2 GeV in terms of only the lowest order $\chi PT$ amplitudes, properly
%  unitarized in terms of the Bethe Salpeter equation, and a cut off to
%  regularize the loops and effectively incorporate effects of the second order
%  Lagrangian [8].
%  
%  The use of the Lipmann Schwinger equation in coupled channels using the
%  information of lowest and second order chiral Lagrangians proved to be an
%  efficient method to deal with the meson baryon interaction [9, 10, 11].
%  The low energy data for $K^- p$ and coupled channels interaction was also well
%  reproduced in [12] using only the lowest order chiral amplitudes and
%  unitarization with the Bethe Salpeter equation plus a cut off. This required
%  however the inclusion of all channels which couple one meson of the ($0^-$)
%  stable octet and one baryon form the ($1/2^+$) stable octet.
%  
%  
%  The particular structure of this resonance in terms of meson baryon
%  components provided by the chiral approach allows one to obtain the coupling
%  of a pion or $\eta$ meson to this resonance and this is the aim of the present
%  paper.
 
%==================================================================
\section{Chiral unitary approach to the $\pi N$ interaction and coupled
channels}
%==================================================================

Let us consider the zero charge state with a meson and a baryon. 
The coupled channels in the $S=0$ sector are: 
$\pi^- p$, $\pi^0 n$, $\eta n$, $K^+\Sigma^-$,
$K^0\Sigma^0$, $K^0\Lambda$. 
The lowest order chiral Lagrangian for the
meson baryon interaction is given by \cite{2,3,4} 
\be
L_1^{(B)} &=& 
\bra \bar{B} i \gamma^{\mu} \nabla_{\mu} B \ket 
- M_B \bra \bar{B} B \ket 
\nonumber \\
&+&
\frac{1}{2} D \bra 
\bar{B} \gamma^{\mu} \gamma_5 \left\{ u_{\mu}, B \right\} \ket
+ \frac{1}{2} F \bra \bar{B} \gamma^{\mu} \gamma_5 [u_{\mu}, B] \ket 
\, , 
\label{LB1}
\ee
where the symbol $< \, >$ denotes the trace of SU(3) matrices and
% \begin{equation}
%       \begin{array}{l}
%               \nabla_{\mu} B = \partial_{\mu} B + [\Gamma_{\mu}, B] \, , \\
%               \Gamma_{\mu} = \frac{1}{2} (u^+ \partial_{\mu} u 
%               + u \partial_{\mu} u^+) \, , \\
%               U = u^2 = {\rm exp} (i \sqrt{2} \Phi / f) \, , \\
%               u_{\mu} = i u ^+ \partial_{\mu} U u^+ \, .
%       \end{array}
% \end{equation}
\be
                \nabla_{\mu} B &=& \partial_{\mu} B + [\Gamma_{\mu}, B] \,
                ,\nonumber \\
                \Gamma_{\mu} &=& \frac{1}{2} (u^+ \partial_{\mu} u 
                + u \partial_{\mu} u^+) \, , \nonumber \\ 
                \label{LB1_exp}
                U &=& u^2 = {\rm exp} (i \sqrt{2} \Phi / f) \, , \\
                u_{\mu} &=& i u ^+ \partial_{\mu} U u^+ \, \nonumber.
\ee
The SU(3) coupling constants which are determined by semileptonic decays of 
hyperons \cite{pr} are 
$F \sim 0.46$, $D \sim 0.79$ ($F+D = g_{A} = 1.26$).  

The SU(3) matrices for the mesons and the baryons are the following
\begin{equation}
\label{matphi}
        \Phi =
        \left(
        \begin{array}{ccc}
                \frac{1}{\sqrt{2}} \pi^0 + \frac{1}{\sqrt{6}} \eta & \pi^+ & K^+ \\
                \pi^- & - \frac{1}{\sqrt{2}} \pi^0 + \frac{1}{\sqrt{6}} \eta & K^0 \\
                K^- & \bar{K}^0 & - \frac{2}{\sqrt{6}} \eta
        \end{array}
        \right) \, ,
\end{equation}
\begin{equation}
\label{matB}
        B =
        \left(
        \begin{array}{ccc}
                \frac{1}{\sqrt{2}} \Sigma^0 + \frac{1}{\sqrt{6}} \Lambda &
                \Sigma^+ & p \\
                \Sigma^- & - \frac{1}{\sqrt{2}} \Sigma^0 
                + \frac{1}{\sqrt{6}} \Lambda & n \\
                \Xi^- & \Xi^0 & - \frac{2}{\sqrt{6}} \Lambda
        \end{array}
        \right) \, .
\end{equation}

In order to describe the $N^*(1535)$ as a meson-baryon resonance state, 
we need interactions of the type $MB \to MB$.  
At lowest order in momentum, that we will keep in our study, the interaction
Lagrangian comes from the $\Gamma_{\mu}$ term in the covariant 
derivative and we find 
\begin{equation}
\label{Lmbmb}
        L_{MB \to MB} = \bra \bar{B} i \gamma^{\mu} \frac{1}{4 f^2}
        [(\Phi \partial_{\mu} \Phi - \partial_{\mu} \Phi \Phi) B
        - B (\Phi \partial_{\mu} \Phi - \partial_{\mu} \Phi \Phi)] \ket \, .  
\end{equation}
This leads to a common structure of the type
$\bar{u} \gamma^u (k_{\mu} + k'_{\mu}) u$ 
for different meson-baryon channels, where 
$u, \bar{u}$ are the Dirac spinors and $k, k'$ the momenta of the incoming
and outgoing mesons. The baryon pole diagrams, which contain two vertices
involving the $D$ and $F$ terms of the Lagrangian in eq. (1), give rise
to a $p$-wave contribution, which we do not consider here, when the positive
energy part of the baryon propagator is taken. In that same case, 
the term from the negative energy
part of the propagator contributes a small amount to the $s$-wave
(of the order of 6 per cent) and we neglect it.

Simple algebraic manipulations of the amplitudes of eq. (\ref{Lmbmb}) 
allow us to write
the lowest order amplitudes as
\begin{equation}
\label{Vij}
        V_{i j} = - C_{i j} \frac{1}{4 f^2} \bar{u} (p') \gamma^{\mu} u (p)
        (k_{\mu} + k'_{\mu}) \, ,
\end{equation}
where $p, p' (k, k')$ are the initial and final momenta of the baryons (mesons)
and $i, j$ indicate any of the channels mentioned above. 
Components of the symmetric matrix $C_{i j}$ are given in Table 1. We are 
interested only in the $s$-wave part of the interaction, in which the 
$N^*(1535)$ is generated, hence the largest contribution comes from
the $\gamma^0$ component of eq. (6). However we keep terms up to 
${\cal O}(k/M)^2$ both
from the $\gamma^0$ and $\gamma^i$ components and this leads 
to a simple expression for the
matrix elements
\begin{equation}
%       V_{i j} = - C_{i j} \frac{1}{4 f^2} (k^0 + k'^0) \, .
V_{i j} = - C_{i j} \frac{1}{4 f^2}(2\sqrt{s} - M_{Bi}-M_{Bj})
\left(\frac{M_{Bi}+E}{2M_{Bi}}\right)^{1/2} \left(\frac{M_{Bj}+E^{\prime}}{2M_{Bj}}
\right)^{1/2}\, ,
\end{equation}
in the meson-baryon center of mass (CM) frame, 
where $M_{Bi}$ and $M_{Bj}$ are the masses
of the baryons $B_{i}$ and $B_{j}$,  and $E$, $E^\prime$ 
the relativistic energies of the incoming ($i$) and
outgoing ($j$) baryons.
 
\begin{table}
\centering
\caption{\small $C_{i j}$ coefficients of eq. (\ref{Vij}). 
$C_{j i} = C_{i j}$.}
\vspace{0.5cm}
\begin{tabular}{c|cccccc}
        & $K^+ \Sigma^-$ & $K^0 \Sigma^0$ & $K^0 \Lambda$ & $\pi^- p$ &
        $\pi^0 n$ & $\eta n$  \\
        \hline
        $K^+ \Sigma^-$ & 1 & $-\sqrt{2}$ & 0 & 0 & $-\frac{1}{\sqrt{2}}$ &
        $-\sqrt{\frac{3}{2}}$  \\
        $K^0 \Sigma^0$ &  & 0 & 0 & $-\frac{1}{\sqrt{2}}$ &
        $-\frac{1}{2}$ & $\frac{\sqrt{3}}{2}$  \\
        $K^0 \Lambda$ &  &  & 0 & $-\sqrt{\frac{3}{2}}$ & $\frac{\sqrt{3}}{2}$ &
        $-\frac{3}{2}$ \\
        $\pi^- p$  &  &  &  & 1 & $-\sqrt{2}$ & 0  \\

        $\pi^0 n$ &  &  &  &  & 0 & 0  \\
        $\eta n$ &  &  &  &  &  & 0 
\end{tabular}
\end{table}

\begin{figure}[tbp]

   \vspace*{1cm}
   \centering
   \footnotesize
   \epsfxsize = 13cm
   \epsfbox{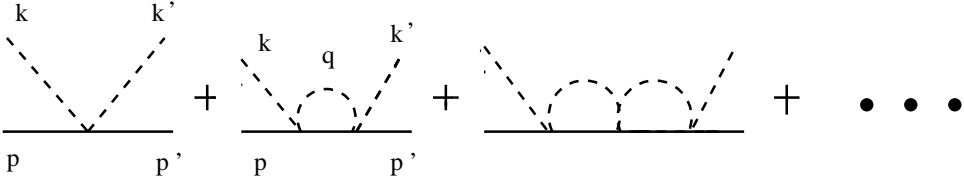} 
% \centerline{\protect
% \hbox{
% \psfig{file=f1BS.eps,height=3.cm,width=13.0cm,angle=-90}}}
   \caption{ \small 
   Diagrammatic representation of the Bethe Salpeter equation.}
   \label{fig_t44}
\end{figure}

Following the steps of Ref. \cite{12} we start from the Bethe Salpeter equation,
depicted diagrammatically in Fig.~\ref{fig_t44},
\begin{equation}
\label{BStij}
        t_{i j} = V_{i j} + V_{i l} \; G_l \; t_{l j}\, ,
\end{equation}
where $t_{ij}$ is the unitary transition matrix from channel $j$ 
to channel $i$ and $G_{l}$ accounts for the intermediate meson 
and baryon propagators for the channel $l$. 
One of the interesting points proved in Ref. \cite{12} is that the $V$ and 
$t$ matrices in that
product appear with their on shell values, where 
the off shell part may be reabsorbed
into renormalization of coupling constants of the lowest order Lagrangian.
This feature was first found in Ref. \cite{8} in the study of the scalar sector of the
meson meson interaction. 
Hence, eq. (\ref{BStij}) reduces to an
algebraic equation where $G$ is given by a diagonal matrix with matrix
elements
\begin{equation}
\label{Gl4int}
        G_{l}(s) = i \, \int_{|\vec{q}\,|< q_{max}} 
        \frac{d^4 q}{(2 \pi)^4} \, \frac{M_l}{E_l
        (\vec{q}\, )} \,
        \frac{1}{k^0 + p^0 - q^0 - E_l (\vec{q}\, ) + i \epsilon} \,
        \frac{1}{q^2 - m^2_l + i \epsilon}\, ,  
\end{equation}
which depends on $p^0 + k^0 =\sqrt{s}$ and $q_{max}$, the cut off which we
choose for $|\vec{q}\,|$ to regularize the loop integral of eq. 
(\ref{Gl4int}).
% Diagrammatically eq. (\ref{Gl4int}) can be written as shown in 
% Fig.~\ref{fig_t44}.

%%   Fig. 1   %%%%%%%%%%%%%%%%%%%%%%%%%%%%%%%%%%%%%%%%%%%%%%%%%%%%%
%%   Fig. 1   %%%%%%%%%%%%%%%%%%%%%%%%%%%%%%%%%%%%%%%%%%%%%%%%%%%%%

The integral in eq. (\ref{Gl4int}) 
without the cut off is logarithmically divergent. 
A subtraction of $G_l(s ')$ for a fixed value of $s '$ makes it convergent
but leads to an undetermined constant. 
The use of different subtraction
constants in each of the channels is one way to account for SU(3) symmetry
breaking, beyond the one provided by the unequal masses of the particles,
and of effectively taking into account effects of higher order Lagrangians
\cite{D}. 
In this way we can then choose an alternative, and equivalent, way to 
account 
for this freedom by still evaluating $G_l$ of eq. (\ref{Gl4int}) 
with a cut off, equal for
all channels and a subtraction constant. 
The cut off is taken larger than the
on shell momenta of any particle in the intermediate states and in practice 
we take the value $q_{max}$ = 1 GeV. 
Hence, after performing analytically the
$q^0$ integration in eq. (\ref{Gl4int}), we find 
\begin{equation} 
\label{Gl3int}
G_{l} = \int_{q_{max}=1\; GeV} 
        \, \frac{d^3 q}{(2 \pi)^3} \, \frac{1}{ 2 \omega_l      (\vec q\, )}
        \,
        \frac{M_l}{E_l (\vec{q}\, )} \,
        \frac{1}{p^0 + k^0 - \omega_l (\vec{q}\, ) - E_l (\vec{q}\, ) 
        + i \epsilon} + a_l \, .
\end{equation}
In eq. (\ref{Gl3int}) one could also have contributions from Castillejo, 
Dalitz,  Dyson (CDD)
poles \cite{CDD}, as they play the role of a dispersion relation. 
In this way one
can account for the role played by resonances corresponding to physical
states which are not generated by the multiple scattering series implicit in
eq. (8).  We are here interested in the $N^*(1535)$ which is generated
dynamically by the series of eq. (8), and other resonances with the same 
quantum numbers will appear at higher energies. The next resonance with the same
quantum numbers is the $N^\ast(1650)$ which lies 115 MeV above. Its contribution
in a narrow region of energies around the $N^\ast(1535)$, where we are 
interested,
can be approximately accommodated by means of suitable subtraction constants in
eq. (10) and this is the point of view that we shall follow there. However, 
one should be cautious when extrapolating the present results at energies
higher than those of the $N^\ast(1535)$ resonance and we shall discuss that
later on.

In order to keep isospin symmetry in the case that the masses of  
the particles
in the same multiplet are equal, we choose $a_l$ 
to be the same for states
belonging to the same isospin multiplet. 
Hence we have four subtraction
constants, $a_{K\Sigma}$, $a_{K\Lambda}$, 
$a_{\pi N}$, $a_{\eta N}$ which
are considered free parameters and are 
fitted to the data. Simultaneously we
take values of meson decay constants $f$ different for $\pi$, $K$ 
or $\eta$ couplings as is the case in $\chi PT$, namely $f_K = 1.22 f_{\pi}$,
$f_\eta=1.3 f_{\pi}$ and $f_{\pi}=93$ MeV \cite{1}.

With all these ingredients we find an acceptable solution to the low energy 
scattering data for the cross sections of 
$\pi^- p\rightarrow\eta n$, 
$\pi^- p\rightarrow K^0\Lambda$, 
$\pi^- p\rightarrow K^0\Sigma^0$, 
and phase shifts and inelasticities of the $I=1/2$ $\pi N$ 
scatterings,  by choosing the parameters
\begin{equation}
        a_{\pi N} = 35\;  {\rm MeV} \, ,   \; \; 
        a_{\eta N} = 16\;  {\rm MeV} \, ,  \; \; 
        a_{K\Lambda} = 40\;  {\rm MeV} \, , 
        a_{K\Sigma}= -21\;  {\rm MeV} \, . \nonumber
\end{equation}

\begin{figure}[tbp]
     \centering
      \footnotesize
      \epsfxsize = 9cm
      \epsfbox{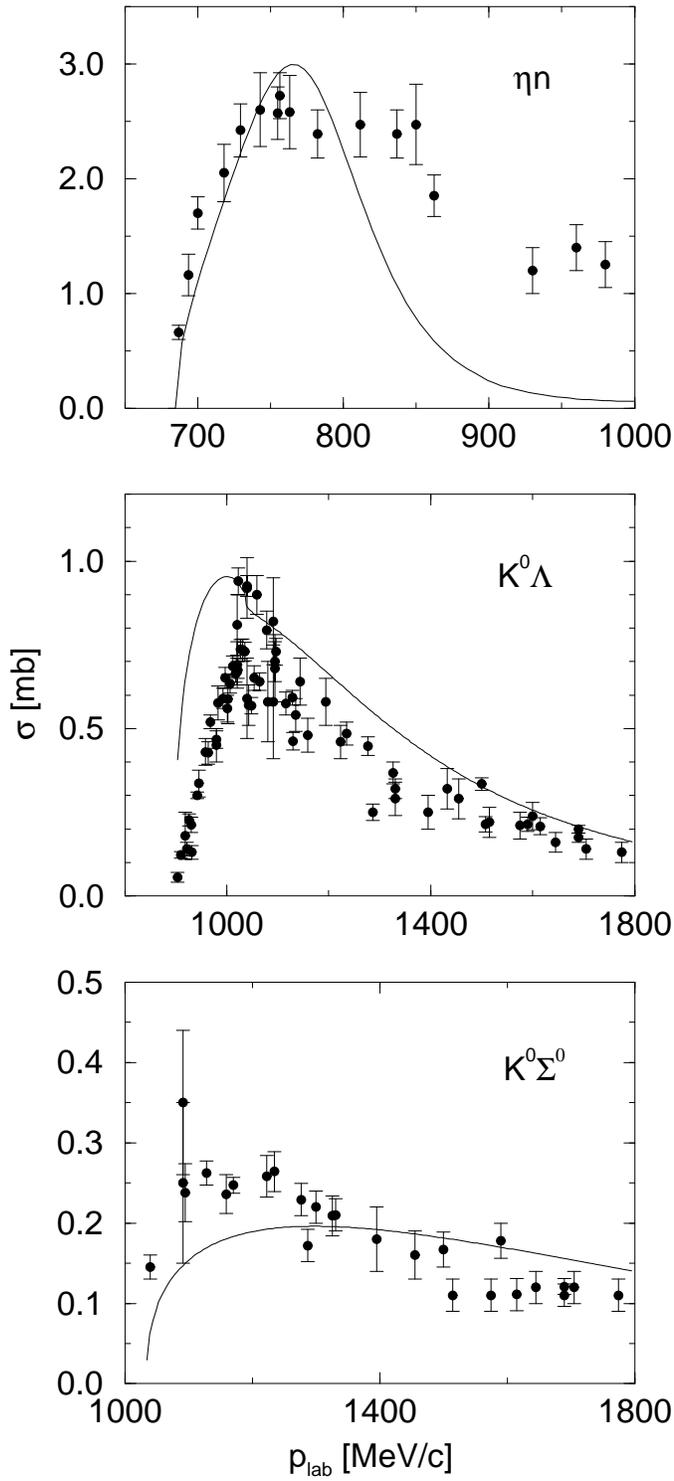} 
%       \centerline{\protect
%       \hbox{
%       \psfig{file=cross.eps,height=10cm,width=13.5cm,angle=-90}}}
        \caption{\small Cross sections for $\pi^- p\rightarrow\eta n, K^0 \Lambda$
and $K^0 \Sigma^0$ reactions as functions of the $\pi^-$ laboratory
momentum.}
        \label{fig_cross}
\end{figure}

\begin{figure}[tbp]
     \centering
      \footnotesize
      \epsfxsize = 9cm
      \epsfbox{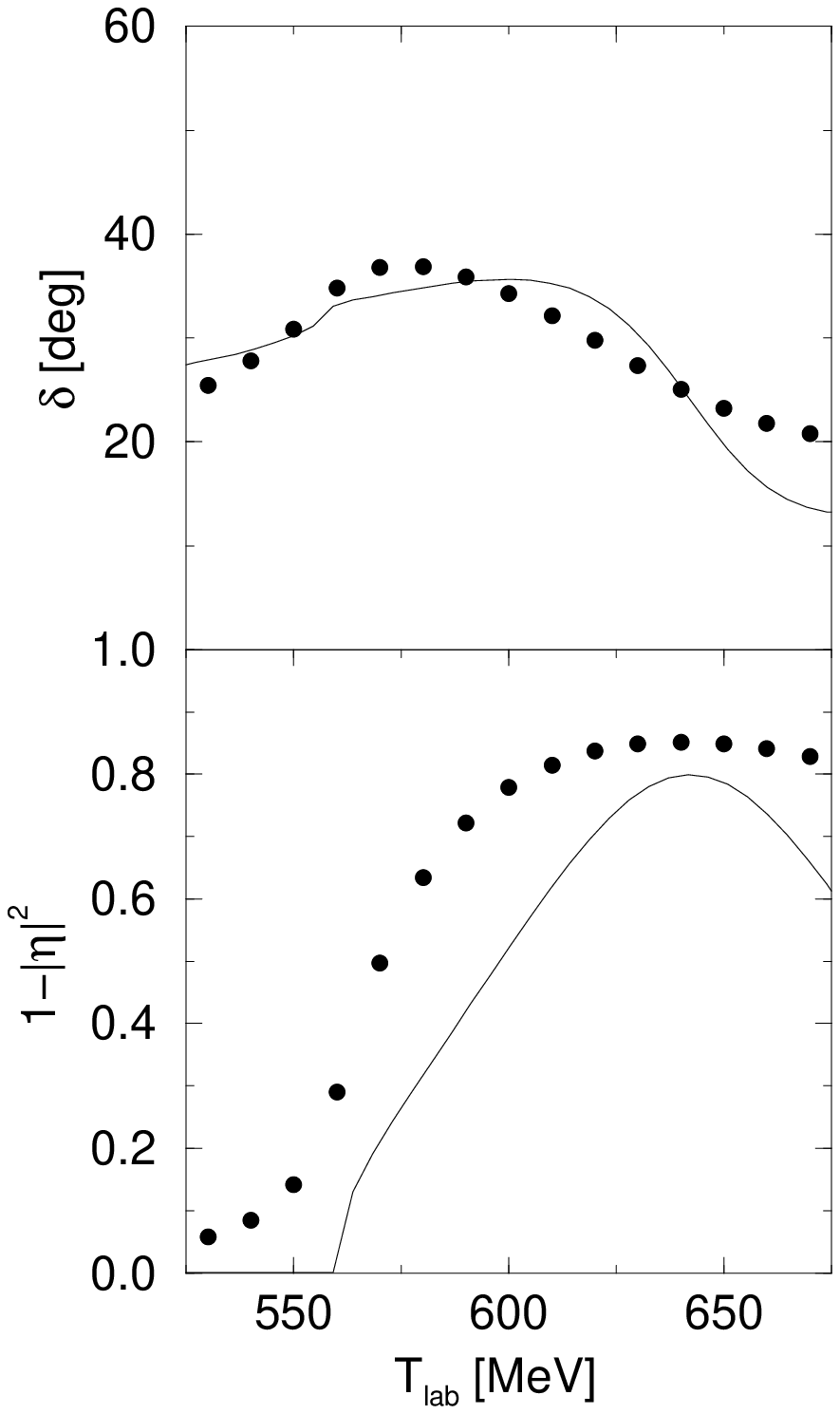} 
%       \centerline{\protect
%       \hbox{
%       \psfig{file=phase.eps,height=10cm,width=13.5cm,angle=-90}}}
        \caption{\small Phase-shifts and inelasticities for 
$\pi N $ scatteering in the isospin $I=1/2$ channel.}
        \label{fig_phase}
\end{figure}

As an example of the quality of the theoretical 
results,  we show the cross sections
for the $\pi^- p\rightarrow\eta n$,
 $\pi^- p\rightarrow K^0\Lambda$,
 $\pi^- p\rightarrow K^0\Sigma^0$
reactions in Fig.~\ref{fig_cross}, and  
the phase shifts and inelasticities in Fig. \ref{fig_phase}. 
The agreement obtained
with the data is fair considering the restrictions in the model, with a reduced
set of parameters. 
The relatively large discrepancies in the inelasticities
can be understood as a lack of the $\pi \pi N$ decay channel, which is visible
already at the threshold of the $\eta N$ channel, where our model starts
producing nonzero values for $1-|\eta^2|$. The results obtained in Ref. \cite{11} 
could be considered slightly better at the price of using a larger
set of free parameters. The model of Ref. \cite{11} has been improved
in a recent work \cite{caro} where
$p$-waves are also included. However, in the latter paper no claims are given
about the $\pi N\rightarrow\pi N$ scattering, which as we can see are fair in
our model for the $I=1/2$ channel, the only one we are interested in here. 
Hence,
no effort has been done to fit the $\pi N$ $I=3/2$ channel, which plays no role
in the present work. We should also note that 
the $\pi^- p\rightarrow K^0\Lambda$ cross section peaks at
around $\sqrt{s} = 1650$ MeV 
and the $\pi^- p\rightarrow K^0\Sigma^0$ one peaks beyond that energy. One should
expect the $N^\ast(1650)$ to play some role there. The fact that one gets
still fair results ignoring the direct coupling of this resonance to the
$K \Lambda$ and $ K \Sigma$ channels would be in consonance with the information
of the Particle Data Table \cite{PDG} which quotes very small couplings of the 
$N^\ast(1650)$ to these channels, as well as to the $\eta N$ one (contrary to the
$N^\ast(1535)$).

%%   Fig. 2   %%%%%%%%%%%%%%%%%%%%%%%%%%%%%%%%%%%%%%%%%%%%%%%%%%%%%
%%   Fig. 2   %%%%%%%%%%%%%%%%%%%%%%%%%%%%%%%%%%%%%%%%%%%%%%%%%%%%%

%%   Fig. 3   %%%%%%%%%%%%%%%%%%%%%%%%%%%%%%%%%%%%%%%%%%%%%%%%%%%%%
%%   Fig. 3   %%%%%%%%%%%%%%%%%%%%%%%%%%%%%%%%%%%%%%%%%%%%%%%%%%%%%

We find the mass of the $N^\ast(1535)$ at an energy around 1550 MeV
from the peak of $|\Im  t_{\eta N\rightarrow \eta N}|$, compatible
with the dispersion of the masses ($1520 - 1555$) MeV quoted
in the PDG \cite{PDG}.

As for the total width and partial decay widths of the resonance
we conduct the following analysis. We observe that both the
$\pi^- p\rightarrow\eta n$ and $\eta n\rightarrow\eta n$ amplitudes
(both in $I=1/2$) are well approximated by the Breit Wigner form 
together with a small background $B(s)$ 
($\sqrt{s}$ in units of MeV):
\begin{equation}
B_{\eta n,\eta n}(s) = \left(-i0.01 \frac{\sqrt{s}-1450}{150}\right) \; 
{\rm MeV}^{-1}\, , \; 
{\rm for} \; 1450 \le \sqrt{s} \le 1650 \; {\rm MeV} \, . 
% 
% \hspace{0.5cm}
% \sqrt{s}\, \hspace{0.3cm} \epsilon\, \hspace{0.3cm}[1450 - 1650]\, {\rm MeV}
\end{equation} 
\begin{equation}
B_{\pi^-p,\eta n}(s) = \left(+0.01 -i 0.025\frac{\sqrt{s}-1650}{200}\right) \; 
{\rm MeV}^{-1}\, , \; 
{\rm for} \; 1450 \le \sqrt{s} \le 1650 \; {\rm MeV} \, . 
% \hspace{0.5cm}
% \sqrt{s}\,\hspace{0.3cm} \epsilon\, \hspace{0.3cm}[1450 - 1650]\, {\rm MeV}
\end{equation} 
Hence we have: 
\begin{equation}
t(\eta n\rightarrow\eta n) = g_\eta^2\frac{1}{\sqrt{s} - M^\ast +
i\frac{\Gamma(s)}{2}} + B_{\eta n,\eta n}(s)
\end{equation}
\begin{equation}
t(\pi^- p\rightarrow\eta n) = g_\pi g_\eta\frac{\sqrt{2}}{\sqrt{s} - M^\ast +
i\frac{\Gamma(s)}{2}} + B_{\pi^-p,\eta n}(s)
\end{equation}
In eqs. (14) and (15)
 the factor $\sqrt{2}$ is an isospin factor and   
$\Gamma(s) = \Gamma_\pi(s) + \Gamma_\eta(s)$, where the partial decay widths
are given, following the nomenclature of Ref.~\cite{chiang}, by:
\begin{equation}
\Gamma_\pi(s)= 6\frac{g_\pi^2}{4\pi}\frac{M}{M^\ast} q_\pi
\end{equation}
\begin{equation}
\Gamma_\eta(s)= 2\frac{g_\eta^2}{4\pi}\frac{M}{M^\ast} q_\eta
\end{equation}
with $q_\pi$, $q_\eta$ the $\pi$, $\eta$ momenta for the decay of a resonance
of mass $\sqrt{s}$ into $\pi N$, $\eta N$ respectively. The values of the
couplings $g_\pi$, $g_\eta$ which lead to a good reproduction of both amplitudes
are

\begin{equation}
g_\pi = \pm 0.55\hspace{0.3cm} g_\eta = \mp 1.73 \ .
\end{equation}
With these values one can see that the size of the background is about
10-15 $\%$ of that of the resonance at the peak of the amplitudes.

The actual partial decay widths of the resonance for $\pi N$ and $\eta N$ 
decay are obtained
by folding $\Gamma_\pi(s)$ and $\Gamma_\eta(s)$ with the mass distribution of
the resonance, which is proportional to the imaginary part of the amplitudes. We
obtain:
\begin{equation}
\Gamma_\pi\simeq 43 {\rm MeV}, \hspace{0.3cm} \Gamma_\eta\simeq 67 {\rm MeV},
\hspace{0.3cm} 
\Gamma_{\pi+\eta} \simeq
110 {\rm MeV}  \ .
\end{equation}
The width and the partial decays 
would be compatible with the dispersion of the data in Ref. \cite{PDG} 
considering that $\pi\pi N$ decay channels are absent in the calculations.
The partial decay widths are compatible with the present data for the
branching ratios, $\Gamma_\pi$ lying on the lower side and $\Gamma_\eta$ on the
upper one.

%==================================================================
\section{Scattering amplitude with $MB^{*}B^{*}$ Couplings}
%==================================================================

%--------------------------
\subsection{General remarks}
%--------------------------

\begin{figure}[tbp]

\begin{center}
\epsfysize = 3cm
\epsfbox{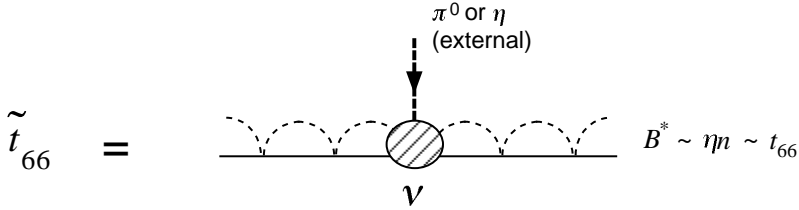}
\caption{\small A diagram for the scattering amplitude of 
the meson($M$)-resonance($B^*$) coupling $\tilde t$.
For $B^* \sim \eta n$, the relevant matrix element is 
$\tilde t_{66}$. }
\label{fig_tildet44}
\end{center}
\end{figure}

\begin{figure}[tbp]
\begin{center}
\epsfysize = 3cm
\epsfbox{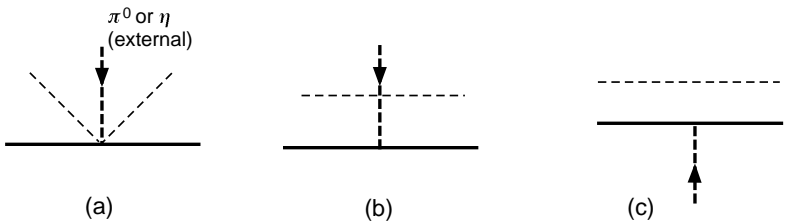}
\caption{\small Diagrammatic representation for the vertex $\calV$ of 
Fig.~\ref{fig_tildet44}.
    Although the diagram (c) is disconnected, after the inclusion of 
        rescattering terms of $t_{66}$, it will become connected as shown in 
        Fig.~\ref{fig_rescatt}(c)}
\label{fig_element}
\end{center}
\end{figure}

Meson-baryon resonance couplings are obtained by inserting an external 
meson line to the resonance state as shown in Fig.~\ref{fig_tildet44}.  
We shall call this meson the probe meson to distinguish it from the
incoming or outgoing mesons which carry the energy to produce the 
resonance.  
The relevant scattering matrix is denoted as $\tilde t_{ij}$.  
As external mesons we take $\pi^{0}$ and $\eta$.  
Furthermore, we specify the resonance state in the channel of 
$\eta n$ which is purely isospin $1/2$ and is largely dominated by  
the $N^\ast(1535)$ pole. Therefore, following the ordering of the states
in Table 1, the corresponding matrix elements are 
$t_{66}$ and $\tilde t_{66}$ for $\eta n\rightarrow\eta n$ and 
$\eta n\rightarrow\pi^0(\eta)\eta n$ respectively.  
There are several different mechanisms for the meson insertion.  
Among them, we need to pick up those which have the 
resonance state both in the initial and final states.  
Therefore, $\tilde t_{66}$ looks like 
\be
\label{mbsbs}
\tilde t_{66} \sim t_{66} \calV t_{66}\, , 
\ee
where $\calV$ represents the elementary vertex functions 
containing the external meson.  
There are three kinds of them as shown in Fig. \ref{fig_element}: 
\begin{description}
        \item[Fig. \ref{fig_element}(a)]  The point like three 
        meson--baryon ($BBMMM$) vertex in which the external meson 
        line is attached at the $BBMM$ vertex of $V_{ij}$.  

        \item[Fig. \ref{fig_element}(b)]  The one connected through a
        meson-meson ($M_1 M_2 \rightarrow M_1 '  M_2 '$ )
    vertex.

        \item[Fig. \ref{fig_element}(c)]  The one in which the external meson 
        couples to the baryon in the meson-baryon propagator $G_l$.  
        Although this diagram itself looks disconnected and irrelevant, 
        after the inclusion of 
        rescattering terms, it will become relevant for $\tilde t_{66}$.  

\end{description}

%%%%%%%%%%%%%%%%%%%%%%%%%%%%%%%%%%%%%%%%%%%%%%%%%%%%%%%%%%%%
%
%   Please comment out the following figure environment
%   because I have made this for OzTeX for Macintosh
%
%%%%%%%%%%%%%%%%%%%%%%%%%%%%%%%%%%%%%%%%%%%%%%%%%%%%%%%%%%%%

%%   Fig. 4   %%%%%%%%%%%%%%%%%%%%%%%%%%%%%%%%%%%%%%%%%%%%%%%%%%%%%
%%   Fig. 4   %%%%%%%%%%%%%%%%%%%%%%%%%%%%%%%%%%%%%%%%%%%%%%%%%%%%%

%%   Fig. 5   %%%%%%%%%%%%%%%%%%%%%%%%%%%%%%%%%%%%%%%%%%%%%%%%%%%%%
%%   Fig. 5   %%%%%%%%%%%%%%%%%%%%%%%%%%%%%%%%%%%%%%%%%%%%%%%%%%%%%

%%   Fig. 6   %%%%%%%%%%%%%%%%%%%%%%%%%%%%%%%%%%%%%%%%%%%%%%%%%%%%%
\begin{figure}[tbp]
\begin{center}
\epsfysize = 3cm
\epsfbox{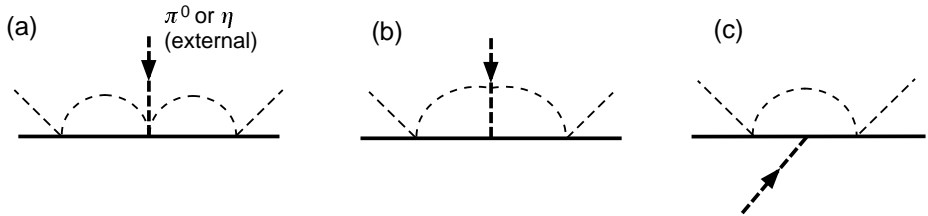}
\caption{\small Vertex functions of $\tilde t_{66}$ to lowest order in the 
rescattering.  They are obtained by attaching $t_{66}$ in the lowest order 
to the 
elementary diagrams given in Fig. \ref{fig_element}. }
\label{fig_rescatt}
\end{center}
\end{figure}
%%   Fig. 6   %%%%%%%%%%%%%%%%%%%%%%%%%%%%%%%%%%%%%%%%%%%%%%%%%%%%%

%%%%%%%%%%%%%%%%%%%%%%%%%%%%%%%%%%%%%%%%%%%%%%%%%%%%%%%%%%%%
%
%   Please recover the following commented out 
%   figure environment you created. 
%
%%%%%%%%%%%%%%%%%%%%%%%%%%%%%%%%%%%%%%%%%%%%%%%%%%%%%%%%%%%%
%
% \vspace{1.5cm}
% \begin{figure}[tbhp]
%       \centerline{\protect
%       \hbox{
%       \psfig{file=f4V0.eps,height=9.5cm,width=12.0cm,angle=-90}}}

%       \caption{Elementary vertex functions for $t_{44}$}
%       \label{fig_element}
%\end{figure}
% 
% \vspace{1.5cm}
% \begin{figure}[tbhp]

% %     \centerline{\protect
% %     \hbox{
% %     \psfig{file=f5V.eps,height=9.5cm,width=12.0cm,angle=-90}}}
%       \caption{Vertex functions for 
%       $\tilde t_{44}$ which contain $t_{44}$ of Fig.~\ref{fig_t44} to 
%       lowest order on 
%       both left and right of the elementary vertices of 
%       Fig.~\ref{fig_element}.}
%       \label{fig_rescatt}
% \end{figure}
%%%%%%%%%%%%%%%%%%%%%%%%%%%%%%%%%%%%%%%%%%%%%%%%%%%%%%%%%%%%

Now by attaching $t_{66}$ to the left and right of $\calV$, we 
obtain $\tilde t_{66}$.  
First we consider the loop structure of $\tilde t_{66}$.  
The corresponding lowest order diagrams are shown in Figs.~\ref{fig_rescatt}(a), 
(b) and (c).   
There are new types of loop integrals.  
We can, however, simplify and relate them 
to that of the meson baryon propagator $G_l$ of eq. (\ref{Gl3int})
as follows:  

\begin{description}
        \item[Fig.~\ref{fig_rescatt}(a)]

        Given the structure of the three meson 
    vertices which are factorized out from the integral, 
    each loop integral of Fig.~\ref{fig_rescatt}(a)
    reduces to that of the meson-baryon propagator $G_l$. 

        \item[Fig.~\ref{fig_rescatt}(b)]
        The loop integrals here look new as they contain two meson propagators 
        and one baryon propagator. 
    However, the amplitudes of the diagrams
    (a) and  (b) 
    can combine in a way that one of the intermediate meson propagators 
    of the diagram (b) cancels with part of the numerator coming from the 
    $MM\rightarrow MM$ amplitude
    yielding the same loop as $G_l$. 
    This will be explained in Fig.~\ref{fig_ppppBB}.  

        \item[Fig.~\ref{fig_rescatt}(c)]
        This diagram also looks new as it contains two baryon propagators and 
        one meson propagator.  
    Yet, we can show that in the case of static probe 
        mesons, which we will 
    adopt here, this loop can be derived from $G_{l}$ by 
    differentiating with respect to a kinematical variable, 
    similarly to what is done in quantum
    electrodynamics to derive the Ward identities.
\end{description}

% After all, 
% in order to generate the $\tilde{t}_{44}$ amplitude which has 
% the double propagator of $N^\ast$, 
% we recall that according to eq. (\ref{BStij}) 
% the $N^\ast$ propagator will appear in the
% series $V + VGV + \cdots $. 
% This means that among the diagrams containing
% the three meson vertices, (a)...(d) of Fig.~\ref{fig_rescatt}, 
% the diagram (d) and those with
% rescattering of $M B\rightarrow M B$ on the right and/or the left
% contribute to the $\tilde{t}_{44}$ amplitude.  
% Other diagrams (a), (b) and (c) are therefore 
% not included in $\tilde t_{44}$.  
% The same can be said for the second set of diagrams
% (e)...(h), 
% where (h) and those with rescattering terms 
% contribute to $\tilde{t}_{44}$. 
%    % This means, as advanced above, that
%    % two loops are at least necessary for this type of diagrams.
% In the case of diagrams (i), (j), (k), all of them and with rescattering  
% terms will contribute to $\tilde{t}_{44}$.

%--------------------------
\subsection{$BBM$ vertex}
%--------------------------
In order to proceed, we investigate the structure of 
various vertices depicted in Figs.~\ref{fig_element} and 
\ref{fig_rescatt} in more detail.  
The simplest are the vertices of the Yukawa type ($BB'M$) coupling.  
They are easily derived from the $D$ and $F$ terms of eq. (\ref{LB1})
expanding $U$ up to one meson field.   
% while the $B B' M_1 M_2 M_3$ vertices
% are obtained from the same terms expanding up to three meson fields. 
We find in a nonrelativistic reduction of the $\gamma^{\mu}\gamma^5$ matrix
\begin{equation}
\label{t_bbm}
        -it_{B' B M} = A_{B' B M} \vec{\sigma}\cdot\vec{q}\, '\, ,  
\end{equation}
with
\begin{equation}
\label{A_bbm}
        A_{B' B M} 
        = \left( x_{B' B M} \frac{D + F}{2f} + y_{B' B M}
        \frac{D - F}{2f} \right) \, ,
\end{equation}
where $\vec{q}\, '$ refers to the momentum of the incoming probe 
meson $M$. 
The coefficients $x_{B' B M}$ and $y_{B' B M}$ are given in Table 2 
for diagonal components. 
We need only diagonal
matrix elements in the baryons, except for the possible case of
$\Sigma^0\Lambda\pi^0$ in which  
$x_{\Sigma^0 \Lambda \pi^0} = \frac{1}{\sqrt{3}}$ and 
        $y_{\Sigma^0 \Lambda \pi^0} = \frac{1}{\sqrt{3}}$.
        We shall also use in eq. (22) the $f$ corresponding to the
        meson which couple to the baryon, $f_\pi$ or $f_\eta$. 

\begin{table}
        \centering
        \footnotesize
        \caption{\small  Diagonal coefficients $x_{B' B M}$ and $y_{B' B M}$ of eq. 
        (\ref{A_bbm}).}.
  \vspace{0.5cm}
        \begin{tabular}{c|ccccc}
                & $\Sigma^-$ & $\Sigma^0$ & $\Lambda$ & $ p$ &
                $ n$   \\
                \hline
                $x_{B B \pi^0}$ & $-1$ & 0& 0 & 1 & $-1$  \\
                $y_{B B \pi^0}$ & 1 & 0 & 0 & 0 & 0 \\
                $x_{B B \eta}$ & $\frac{1}{\sqrt{3}}$  & $\frac{1}{\sqrt{3}}$ 
                & $-\frac{1}{\sqrt{3}}$  & $\frac{1}{\sqrt{3}}$ & $\frac{1}{\sqrt{3}}$  \\
                $y_{B B \eta}$  & $\frac{1}{\sqrt{3}}$  & $\frac{1}{\sqrt{3}}$ 
                & $-\frac{1}{\sqrt{3}}$ & $-\frac{2}{\sqrt{3}}$ & $-\frac{2}{\sqrt{3}}$   
        \end{tabular}
\end{table}

%--------------------------
\subsection{$BBMMM$ vertex}
%--------------------------
This vertex appears in the diagrams depicted in Figs.~\ref{fig_element}(a) and 
\ref{fig_rescatt}(a).  
The Lagrangians for the three meson--baryon 
vertices are obtained expanding $u_\mu$
in the $F$, $D$ terms of eq. (\ref{LB1}) up to three meson fields. 
We obtain 
\begin{equation}
\label{u_3m}
u_{\mu} = \frac{\sqrt{2}}{12f^3} (\partial_\mu \Phi\Phi^2 - 2\Phi\partial_\mu
\Phi\Phi + \Phi^2\partial_\mu\Phi)\, .
\end{equation}
Expressions for the matrix elements of $u_\mu$ involving only nucleons were
obtained in Ref.~\cite{meiss}. Here we have to generalize it for the case of hyperons also.
Once again, by taking the nonrelativistic approximation of the $\gamma$
matrices, $\gamma^{\mu}\gamma^5\rightarrow\sigma^i$, we obtain
\begin{equation}
\label{t_bbmmm}
-it_{\alpha ' \alpha M} = C_{\alpha' \alpha M} \vec{\sigma}\cdot\vec{q}\, '
\, , 
\end{equation}
with 
\begin{equation}
\label{C_bbmmm}
C_{\alpha ' \alpha M} 
= \frac{1}{12 f^2} \left( X_{\alpha ' \alpha M} \frac{D + F}{2f}
+ Y_{\alpha ' \alpha M} \frac{D - F}{2f}\right) \, ,
\end{equation}
where the coefficients, $X_{\alpha ' \alpha M}$, $Y_{\alpha ' \alpha M}$, 
with $\alpha$, $\alpha\, '$ indicating a $M B $ state, are given in Tables
3 and 4 
for $M\equiv{\pi}^0$, $\eta$ respectively. Once again the $f$ inside the
brackets refers to $\pi$ or $\eta$, while the $f^2$ in the factor $12f^2$ will become
$f_\alpha$ $f_{\alpha^\prime}$.

%%%%%%%%%%%%%%%%%%%%%%%%%%%%%%%%%%%%%%%%%%%%%%%%%%%%%%%%%%%%%%%%%%%%%%%%%%%%%
%%  BEGIN of Tables for coefficients X and Y's

\begin{table}
        \centering
        \footnotesize
        \caption{\small $X_{lm}$ and $Y_{lm}$ coefficients for the three meson vertex, eq. 
        (25) for $\pi^0$
        coupling. They are symmetric as $X_{l m} = X_{m l}$ and $Y_{l m} = Y_{m l}$.}
        \vspace{0.5cm}
        \begin{tabular}{c|cccccc|cccccc}
       & \multicolumn{6}{ c|}{$X_{lm}$} &
         \multicolumn{6}{ c}{$Y_{lm}$} \\
        \hline
                & $K^+ \Sigma^-$ & $K^0 \Sigma^0$ & $K^0 \Lambda$ & $\pi^- p$ &
                $\pi^0 n$ & $\eta n$  
                & $K^+ \Sigma^-$ & $K^0 \Sigma^0$ & $K^0 \Lambda$ & $\pi^- p$ &
                $\pi^0 n$ & $\eta n$ \\
                \hline
                $K^+ \Sigma^-$ & 0 & 0 & 0 & 0 & 0 & 0  & 
                     $-2$ & 0 & 0 & 0 & $\frac{1}{\sqrt{2}}$ & $\sqrt{\frac{3}{2}}$\\
                $K^0 \Sigma^0$ &  & 1 & $-\frac{1}{\sqrt{3}}$  & 0 & 0 & 0 
                         &  & 1 & $-\frac{1}{\sqrt{3}}$  & $-\frac{3}{\sqrt{2}}$ & 
                         $-\frac{1}{2}$ & $\frac{\sqrt{3}}{2}$ \\
                $K^0 \Lambda$ &  &  & $-1$ & $\sqrt{6}$ & $-\frac{1}{\sqrt{3}}$ & 1 
             &  &  & $-1$ & $-\sqrt{\frac{3}{2}}$ & $\frac{1}{2\sqrt{3}}$ &
             $-\frac{1}{2}$ \\
        $\pi^- p$  &  &  &  & $-4$ & $2\sqrt{2}$ & 0 &  &  &  & 0 & 0 & 0  \\
                $\pi^0 n$ &  &  &  &  & 0 & 0  &  &  &  &  & 0 & 0   \\
                $\eta n$ &  &  &  &  &  & 0  &  &  &  &  &  & 0 
        \end{tabular}
\end{table}
% \begin{table}
%       \centering
%       \footnotesize
%       \caption{\small $Y_{lm}$ coefficients for the three meson vertex, eq. 
%       (19) for $\pi^0$
%       coupling. $Y_{l m} = Y_{m l}$.}
%       \vspace{0.5cm}
%       \begin{tabular}{c|cccccc}
%               & $K^+ \Sigma^-$ & $K^0 \Sigma^0$ & $K^0 \Lambda$ & $\pi^- p$ &
%               $\pi^0 n$ & $\eta n$  \\
%               \hline
%               $K^+ \Sigma^-$ & $-2$ & 0 & 0 & 0 & $\frac{1}{\sqrt{2}}$ & 
%                  $\sqrt{\frac{3}{2}}$  \\
%               $K^0 \Sigma^0$ &  & 1 & $-\frac{1}{\sqrt{3}}$  & $-\frac{3}{\sqrt{2}}$ &
%               $-\frac{1}{2}$ & $\frac{\sqrt{3}}{2}$  \\
%               $K^0 \Lambda$ &  &  & $-1$ & $-\sqrt{\frac{3}{2}}$ & $\frac{1}{2\sqrt{3}}$ &
%               $-\frac{1}{2}$ \\
%               $\pi^- p$  &  &  &  & 0 & 0 & 0  \\
%               $\pi^0 n$ &  &  &  &  & 0 & 0  \\
%               $\eta n$ &  &  &  &  &  & 0 
%       \end{tabular}
% \end{table}
\begin{table}
        \centering
        \footnotesize
        \caption{\small $X_{lm}$ and $Y_{lm}$ coefficients for the three 
        meson vertex, eq.~(25) for $\eta$
        coupling. They are symmetric as $X_{l m} = X_{m l}$ and $Y_{l m} = Y_{m l}$.}
        \vspace{0.5cm}
        \begin{tabular}{c|cccccc|cccccc}
       & \multicolumn{6}{ c|}{$X_{lm}$} &
         \multicolumn{6}{ c}{$Y_{lm}$} \\
        \hline
                & $K^+ \Sigma^-$ & $K^0 \Sigma^0$ & $K^0 \Lambda$ & $\pi^- p$ &
                $\pi^0 n$ & $\eta n$  
                & $K^+ \Sigma^-$ & $K^0 \Sigma^0$ & $K^0 \Lambda$ & $\pi^- p$ &
                $\pi^0 n$ & $\eta n$\\
                \hline
                $K^+ \Sigma^-$ & 0 & $-\sqrt{3}$ & $-1$& 0 & 0 & 0 
                & $-2\sqrt{3}$ & $\sqrt{3}$ & $-1$& 0 & $\sqrt{\frac{3}{2}}$ &
                $\frac{3}{\sqrt{2}}$\\
                $K^0 \Sigma^0$ &  & $-\sqrt{3}$ & 1 & 0 &
                0 & 0  
                 & &$-\sqrt{3}$  & 1 & $\sqrt{\frac{3}{2}}$ & $\frac{\sqrt{3}}{2}$&
                $-\frac{3}{2}$   \\
                $K^0 \Lambda$ &  &  & $\sqrt{3}$ & $-\sqrt{2}$ & 1 &
                $-\sqrt{3}$ 
                 &  &  & $\sqrt{3}$ & $\frac{1}{\sqrt{2}}$ & $-\frac{1}{2}$ &
                $\frac{\sqrt{3}}{2}$ \\
                $\pi^- p$  &  &  &  & 0 & 0 & 0    &  &  &  & 0 & 0 & 0  \\
                $\pi^0 n$ &  &  &  &  & 0 & 0   &  &  &  &  & 0 & 0  \\
                $\eta n$ &  &  &  &  &  & 0  &  &  &  &  &  & 0 
        \end{tabular}
\end{table}
% \begin{table}
%       \centering
%       \footnotesize
%       \caption{\small $Y_{lm}$ coefficients for the three meson vertex, 
%       eq. (19) for $\eta$
%       coupling. $Y_{l m} = Y_{m l}$.}
%       \vspace{0.5cm}
%       \begin{tabular}{c|cccccc}
%               & $K^+ \Sigma^-$ & $K^0 \Sigma^0$ & $K^0 \Lambda$ & $\pi^- p$ &
%               $\pi^0 n$ & $\eta n$  \\
%               \hline
%               $K^+ \Sigma^-$ & $-2\sqrt{3}$ & $\sqrt{3}$ & $-1$& 0 & $\sqrt{\frac{3}{2}}$ &
%               $\frac{3}{\sqrt{2}}$ \\
%               $K^0 \Sigma^0$ & &$-\sqrt{3}$  & 1 & $\sqrt{\frac{3}{2}}$ & $\frac{\sqrt{3}}{2}$&
%               $-\frac{3}{2}$   \\
%               $K^0 \Lambda$ &  &  & $\sqrt{3}$ & $\frac{1}{\sqrt{2}}$ & $-\frac{1}{2}$ &
%               $\frac{\sqrt{3}}{2}$ \\
%               $\pi^- p$  &  &  &  & 0 & 0 & 0  \\
%               $\pi^0 n$ &  &  &  &  & 0 & 0  \\
%               $\eta n$ &  &  &  &  &  & 0 
%       \end{tabular}
% \end{table}

%%  END of Tables for coefficients X and Y's
%%%%%%%%%%%%%%%%%%%%%%%%%%%%%%%%%%%%%%%%%%%%%%%%%%%%%%%%%%%%%%%%%%%%%%%%%%%%%
In eqs. (\ref{t_bbmmm}) and (\ref{C_bbmmm}), 
we have kept only the terms which survive when $t_{66}$ is 
attached to both sides of the three meson vertex, since 
the $N^{*}$ state should be dynamically generated by the 
scattering of the meson.  
Note that the successive $BBMM$-vertices $V_{ij}$, where
the incoming or outgoing meson is scattered,  
are $s$-wave couplings.   
Due to this $s$-wave nature, the terms in eq. (\ref{u_3m}) which involve
the derivatives acting on the two internal mesons 
vanish in the loop integral and only
the term which has the derivative in the external $\pi^0$ or $\eta$ field
survives. 
% As we will see below we will indeed need loops
% to the righ and the left in order to generate the $N^*$ resonance 
% to the right and left of the external $\pi^0$ ($\eta$) coupling.

%--------------------------
\subsection{$MMMM$ vertex}
%--------------------------
Finally we consider the terms involving the meson meson interaction 
which appear in Figs.~\ref{fig_element}(b) and \ref{fig_rescatt}(b). 
The evaluation of the $M_1 M_2 \rightarrow M_1 ' M_2 '$ vertices 
has been done in Ref. \cite{8}.
Here we need in addition the meson propagator and a $B' B M$ vertex in order
to evaluate the diagrams of Fig.~\ref{fig_rescatt}(b) and 
the related ones obtained by
iterating to the right and left the meson baryon scattering potential,
$V_{ij}$. 
Let us look in detail at these diagrams in some cases in order to see their
general behavior. 
%%   Fig. 7   %%%%%%%%%%%%%%%%%%%%%%%%%%%%%%%%%%%%%%%%%%%%%%%%%%%%%
\begin{figure}[tbhp]
   \centering
        \footnotesize
   \epsfxsize = 4cm
   \epsfbox{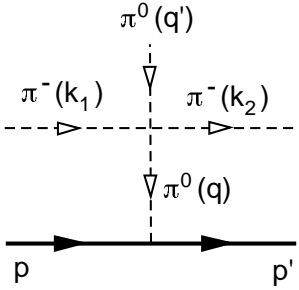} 
%       \centerline{\protect
%       \hbox{
%       \psfig{file=f6pppp.eps,height=5.5cm,width=5.cm,angle=-90}}}
        \caption{\small The case involving the 
		$\pi\pi\rightarrow\pi\pi$ vertex. }
        \label{fig_fourpi}

\end{figure}
%%   Fig. 7   %%%%%%%%%%%%%%%%%%%%%%%%%%%%%%%%%%%%%%%%%%%%%%%%%%%%%

\begin{figure}[tbhp]
   \centering
        \footnotesize
   \epsfxsize = 14cm
   \epsfbox{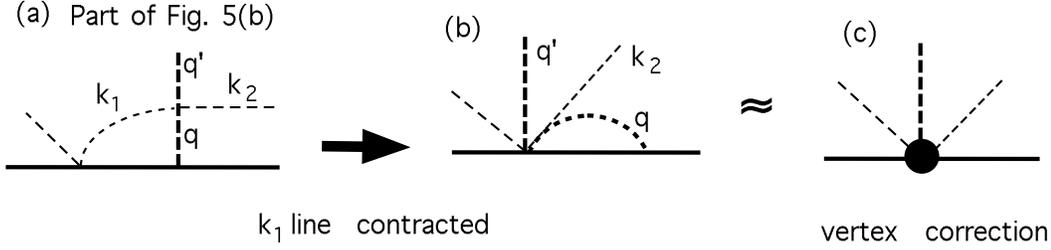} 
%       \centerline{\protect
%       \hbox{
%       \psfig{file=f7cont.eps,height=3.5cm,width=5.cm,angle=-90}}}
        \caption{\small Diagram obtained from the diagram of 
		Fig.~\ref{fig_rescatt}(b)
        when the off shell part
        of the meson meson amplitude is considered. }
        \label{fig_ppppBB}
\end{figure}

First let us take the case involving the $\pi\pi\rightarrow\pi\pi$
vertex as shown in Fig.~\ref{fig_fourpi}.
The contribution of this diagram is:
\begin{equation}
\label{tmmmm_1}
        -it= (-i)\,  t_{\pi^-\pi^0\rightarrow\pi^-\pi^0}\, 
        \frac{i}{q^2 - m_{\pi}^2}\, 
        \frac{D + F}{2f}\, \vec{\sigma}\cdot{\vec{q}}\, ,
\end{equation}
where
\begin{equation}
\label{tmmmm_2}
        t_{\pi^-\pi^0\rightarrow\pi^-\pi^0} 
        = -\frac{1}{3 f^2}[3(k_1 - k_2)^2
        -\sum p_i^2 + m_{\pi}^2]\, .
\end{equation}
We note that the $\pi^-\pi^0\rightarrow\pi^-\pi^0$ 
vertex function has an off
shell behavior since it involves $\sum{p_i^2}$ where $p_i$ 
is the momentum
of each of the pion lines. 
The first step is to recall that, following the steps
of Ref. \cite{8}, the $M M$ amplitude in the loops has to be taken 
with $p_i^2 = m_i^2$,
since the terms with $p_i^2 - m_i^2$ will kill one of 
the off shell propagators
and renormalize other diagrams already considered. 
This is illustrated in Fig. \ref{fig_ppppBB}, where starting from diagram (a), 
the term $k_{1}^2 -m_{\pi}^{2}$  
would eliminate the internal meson propagator of momentum $k_{1}$ 
and lead to
diagram (b), which can be interpreted as 
a vertex correction of the three pion
vertex, i.e., the diagram (a) of Fig. \ref{fig_element}, as shown 
in the diagram Fig.~\ref{fig_ppppBB}(c). 
This can be reabsorbed into the physical coupling of
the vertex of Fig.~\ref{fig_element}(a).

% For instance, the term with
% $k_1^2 - m_\pi^2$ would eliminate the internal 
% meson propagator of diagram
% ($f$) of Fig.~\ref{fig_rescatt} leading to the diagram of 
% Fig.~\ref{fig_ppppBB}.
% This is a vertex correction to the diagram (a) of fig. 4 

%%   Fig. 8   %%%%%%%%%%%%%%%%%%%%%%%%%%%%%%%%%%%%%%%%%%%%%%%%%%%%%
\vspace{1.5cm}
%%   Fig. 8   %%%%%%%%%%%%%%%%%%%%%%%%%%%%%%%%%%%%%%%%%%%%%%%%%%%%%

\vspace{-0.5cm}
Next we consider the first term of eq. (\ref{tmmmm_2})
\begin{equation}
\label{cond1}
(k_1 - k_2)^2 = (q - q ')^2\, .
\end{equation}
In order to have the same kinematics of the $N^\ast$ to the right and left
of the $\pi^0 N^\ast N^\ast$ vertex we shall take $q^{\prime\,{\mu}}\rightarrow 0$
(soft pion limit). We are concerned about the coefficient which will accompany
$\vec{\sigma}\cdot\vec{q}\, ' $ for this vertex and this coefficient will be
roughly independent of $q '$ as is the case of the $\pi N N $ vertex in the 
nonrelativistic limit that we take. 
Hence
\begin{equation}
\label{cond2}
(k_1 - k_2)^2\simeq q^2
\end{equation}
in this limit. Keeping the other terms in eq. (\ref{cond1})
 would lead to $q^2 + m_{\pi}^2$
since the linear term in $q$ would vanish in the loops. The limit taken
neglects $m_\pi^2$ versus $q^2$ which is certainly a fair approximation.
Consistency with that approximation will require the neglect of the terms
proportional to $m_\pi^2$ when they appear and we shall do so in the
evaluations here.

On the other hand the term  $\vec{\sigma}\cdot\vec{q}$ 
in eq. (\ref{tmmmm_1}) can be
rewritten as:
\begin{equation}
\vec{\sigma}\cdot\vec{q} = \vec{\sigma}\cdot(\vec{k}_1 + \vec{q}\, ' -
\vec{k}_2)
\end{equation}
and the terms proportional to $\vec{k}_1$, $\vec{k}_2$ will vanish in the
loops.
 Hence we can write the contribution of the diagram of 
 Fig.~\ref{fig_fourpi} in an effective
 way 
 as:
\begin{equation}
-it = - \frac{1}{f^2}\, (q^2 - m_{\pi}^2)\, \frac{1}{q^2 - m_{\pi}^2}\, 
\frac{D + F}{2f}\vec{\sigma}\cdot{\vec{q}\, '}\, .
\end{equation}

In this equation we see that the meson propagator cancels with the remnant of
the $\pi\pi\rightarrow\pi\pi$ vertex and we obtain a structure of the
type of the three meson vertex of Fig.~\ref{fig_rescatt}(a). 
Chiral symmetry brings always
these two kind of diagrams together as is well known from early studies
of $\pi N\rightarrow\pi\pi N$ interaction \cite{wein} prior to the more systematic $\chi
PT$ approach to these amplitudes.
The result obtained indicates that the contribution of these terms can be cast
in the same way as in eq. (\ref{C_bbmmm}).

%%   Fig. 9   %%%%%%%%%%%%%%%%%%%%%%%%%%%%%%%%%%%%%%%%%%%%%%%%%%%%%
\begin{figure}[tbhp]
   \centering
        \footnotesize
   \epsfxsize = 4cm
   \epsfbox{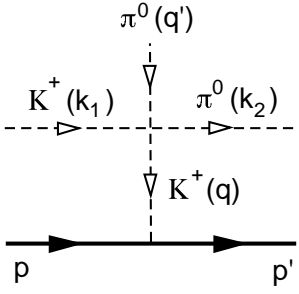} 
%       \centerline{\protect
%       \hbox{
%       \psfig{file=f8ppkk.eps,height=5.5cm,width=5.cm,angle=-90}}}
        \caption{\small Feynman diagram for 
        the $K^+\pi^0\rightarrow K^+\pi^0$ vertex. }
        \label{fig_kkppBB}
\end{figure}
%%   Fig. 9   %%%%%%%%%%%%%%%%%%%%%%%%%%%%%%%%%%%%%%%%%%%%%%%%%%%%%

Let us see another example shown in Fig.~\ref{fig_kkppBB}.
In this case the $K^+\pi^0\rightarrow K^+\pi^0$ vertex, which we take from
the $K^+ K^-\rightarrow \pi^0 \pi^0$ amplitude of Ref. \cite{8} via crossing, can be written
after setting $p_i^2 = m_i^2$ as:
\begin{equation}
-\frac{3}{4f^2}t = -\frac{3}{4f^2}(k_1 - q)^2 = -\frac{3}{4f^2}(k_2 - q ' )^2
\simeq -\frac{3}{4f^2}k_2^2\, ,
\end{equation}
and the on shell condition will rend this as a term proportional to $m_{\pi}^2$
which we will neglect. Note that in addition we would have one loop with two
heavy meson propagators which would further reduce the strength of the diagram.

The argument above would lead to a term proportional to $m_{\eta}^2$ if the
$\pi^0$ with momentum $k_2$ is substituted by an $\eta$. This term would not be
small in principle. However, the $K^+$ attached to the baryon would involve the
$K^+\Sigma^-n$ vertex which goes like ($F - D$), small compared to the $F + D$
component. Furthermore it would contain two loops involving each two heavy
mesons which would further reduce the contribution.

The arguments used above have been used in order to justify that terms of the
type of Fig.~\ref{fig_kkppBB} 
with a heavy meson exchange attached to the baryon line, with one
loop to the right and another one to the left, should be small compared to
those
where just a pion is attached to the baryon line. We will 
just adopt this approximation
and keep only the terms where a pion is exchanged there. 
Under these
approximations the diagrams of the type of Figs.~\ref{fig_fourpi} 
and \ref{fig_kkppBB} can be cast in terms of
an effective operator similar to eq. (\ref{C_bbmmm}) as:
\begin{equation}
-it_{\alpha '\alpha M} = \tilde C_{\alpha '\alpha M} 
\vec{\sigma}\cdot{\vec{q}\, '} \, , 
\end{equation}
where
\begin{equation}
\tilde{C}_{\alpha '\alpha M} 
= \frac{1}{12f^2} \left( \tilde{X}_{\alpha '\alpha M}\frac{D + F}{2f} 
+ \tilde{Y}_{\alpha '\alpha M}\frac{D - F}{2f}\right) \, ,
\end{equation}
and the $\tilde{X}$ and $\tilde{Y}$ coefficients are given in 
Table 5 for an external $\pi^0$
coupling. The same comments given after eq. (25) concerning
the values of the $f^\prime s$ are in order here. For the case of $\eta$ external coupling and intermediate $\pi$
attached to the baryon line, the $M M\rightarrow M M$ vertices with three
pions and one $\eta$ are either zero or proportional to $m_\pi^2$, which we
omit in our analysis. Surviving terms would involve loops with two kaons and
$\Sigma$ or $\Lambda$ which should be suppressed. We will omit these terms
and then within this approximation the $\tilde{X}$, $\tilde{Y}$ coefficients for the
external $\eta$ will be taken zero. In the case of the $\eta$ the soft meson
limit is also a drastic approximation. Altogether this means that for the $\eta
N^\ast N^\ast$ coupling we should admit larger uncertainties than for the
$\pi^0 N^\ast N^\ast$ coupling, where the approximations done are rather sensible.

\begin{table}
        \centering
        \footnotesize
        \caption{\small $\tilde X_{lm}$ and $\tilde Y_{lm}$ 
        coefficients of eq. (34). 
        They are symmetric as $\tilde X_{l m} = \tilde X_{m l}$ 
        and $\tilde Y_{l m} = \tilde Y_{m l}$.}
        \vspace{0.5cm}
        \begin{tabular}{c|cccccc|cccccc}
       & \multicolumn{6}{ c|}{$\tilde X_{lm}$} &
         \multicolumn{6}{ c}{$\tilde Y_{lm}$} \\
        \hline
                & $K^+ \Sigma^-$ & $K^0 \Sigma^0$ & $K^0 \Lambda$ & $\pi^- p$ &
                $\pi^0 n$ & $\eta n$  
                & $K^+ \Sigma^-$ & $K^0 \Sigma^0$ & $K^0 \Lambda$ & $\pi^- p$ &
                $\pi^0 n$ & $\eta n$  \\
                \hline
                $K^+ \Sigma^-$ & 3 & 0 & 0 & 0 & 0 &
                $ 0 $ 
                & $-3$ & 0 & 0 & 0 & 0 &
                $ 0 $ \\
                $K^0 \Sigma^0$ & & 0 & $-\sqrt{3}$ & 0 & 0 & 0   
                 & & 0 & $-\sqrt{3}$ & 0 & 0 & 0   \\
                $K^0 \Lambda$ &  &  & 0 & 0 & 0 &
                $ 0 $ & & & 0 & 0 & 0 & 0 \\
                $\pi^- p$  &  &  &  & $-12$ & 0 & 0  &  &  &  & 0 & 0 & 0  \\
                $\pi^0 n$ &  &  &  &  & 0 & 0 &  &  &  &  & 0 & 0  \\
                $\eta n$ &  &  &  &  &  & 0  &  &  &  &  &  & 0 
        \end{tabular}
\end{table}

% \begin{table}
%       \centering
%       \footnotesize
%       \caption{\small \~Y$_{l m}$ coefficients of eq. (28). \~Y$_{m l}$ = \~Y$_{l m}$.}
% %     \vspace{0.5cm}
%       \begin{tabular}{c|cccccc}
%               & $K^+ \Sigma^-$ & $K^0 \Sigma^0$ & $K^0 \Lambda$ & $\pi^- p$ &
%               $\pi^0 n$ & $\eta n$  \\
%               \hline
%               $K^+ \Sigma^-$ & $-3$ & 0 & 0 & 0 & 0 &
%               $ 0 $ \\
%               $K^0 \Sigma^0$ & & 0 & $-\sqrt{3}$ & 0 & 0 & 0   \\
%               $K^0 \Lambda$ &  &  & 0 & 0 & 0 &
%               $ 0 $ \\
%               $\pi^- p$  &  &  &  & 0 & 0 & 0  \\
%               $\pi^0 n$ &  &  &  &  & 0 & 0  \\
%               $\eta n$ &  &  &  &  &  & 0 
%       \end{tabular}
% \end{table}

To end this subsection, we  note that in diagrams (a), (b) of 
Fig.~\ref{fig_rescatt} we
 can sum the two contributions of the type of Fig. \ref{fig_element} (a),
(b) in one block by means of
\begin{equation}
-t_{\alpha ' \alpha ' M}= D_{\alpha ' \alpha M} \vec{\sigma}\cdot\vec{q}\, '\, ,
\end{equation}
with
\begin{equation}
\label{Dllm}
D_{\alpha ' \alpha M} = \frac{1}{12f^2}\left( (X_{\alpha '\alpha M} 
+  \tilde X_{\alpha ' \alpha M})\frac{D + F}{2f} 
+(Y_{\alpha ' \alpha M} +  \tilde Y_{\alpha ' \alpha M})
\frac{D - F}{2f}\right) \  ,
\end{equation}
with the same considerations about the $f$ factors as given after eq. (25).
%==================================================================
\section{Generation of the $\pi^0(\eta) N^\ast N^\ast$ vertex.}
%==================================================================

\begin{figure}[tbhp]
   \centering
        \footnotesize

   \epsfxsize = 5cm
   \epsfbox{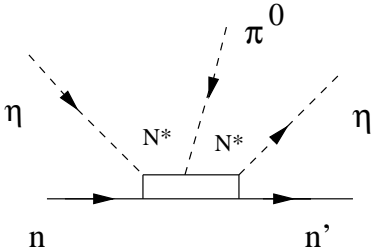} 
%       \centerline{\protect
%       \hbox{
%       \psfig{file=f9nspns.eps,height=5.5cm,width=5.cm,angle=-90}}}
        \caption{\small Diagrammatic representation of the $\pi N^\ast N^\ast$ couplings
        with explicit $N^\ast$ propagators.}
        \label{fig_nspns}
\end{figure}

   From the series of Fig.~\ref{fig_tildet44} 
we must pick up the terms which factorize the $\eta n$
amplitude to the left and the right of the external $\pi^0(\eta)$ vertex
since this amplitude contains the $N^\ast$ pole and hence the resulting
amplitude can be compared to the conventional one which appears from the
diagram of Fig.~\ref{fig_nspns}.  

%%   Fig. 10   %%%%%%%%%%%%%%%%%%%%%%%%%%%%%%%%%%%%%%%%%%%%%%%%%%%%%
%%   Fig. 10   %%%%%%%%%%%%%%%%%%%%%%%%%%%%%%%%%%%%%%%%%%%%%%%%%%%%%

\begin{figure}[tbhp]
   \centering
        \footnotesize
   \epsfxsize = 5cm
   \epsfbox{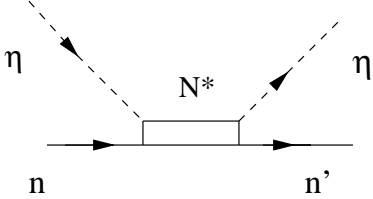} 
%       \centerline{\protect
%       \hbox{
%       \psfig{file=f10ns.eps,height=5.5cm,width=5.cm,angle=-90}}}
        \caption{\small Diagrammatic representation of the $\eta  n\rightarrow\eta n$
        process
        through explicit $N^\ast$ excitation.}
        \label{fig_ns}
\end{figure}

Hence we shall start from $\eta n$ in the series of 
Fig.~\ref{fig_tildet44} and finish up with
$\eta n$ for simplicity. The resulting amplitude will be compared to the one
of Fig.~\ref{fig_nspns} 
which is given by (the index 6 stands for the 6th channel of $\eta n$ in 
Table 1)
\begin{equation}
\label{tm66param}
-i\tilde{t}_{66}(\sqrt{s}) = (-ig_\eta)\frac{i}{\sqrt{s} - M^\ast +
i\frac{\Gamma(s)}{2}}\, C\, \vec{\sigma}\cdot\vec{q}\, '
\frac{i}{\sqrt{s} - M^\ast
+i\frac{\Gamma(s)}{2}}\, (-ig_\eta)\, ,
\end{equation}
where $C\vec{\sigma}\cdot\vec{q}\, '$ is the $N^\ast N^\ast\pi^0$ coupling, 
to be
compared with $- (D + F)\vec{\sigma}\cdot\vec{q}\, '/2f$ of $\pi^0 n n$
in the case of nucleons. 
On the other hand the $\eta n\rightarrow\eta n$ amplitude proceeding
through $N^\ast$ excitation is depicted in Fig.~\ref{fig_ns} and given by
\begin{equation}
\label{t66param}
-it_{66}(\sqrt{s}) = (-ig_\eta)\frac{i}{\sqrt{s} - 
M^\ast+ i\frac{\Gamma(s)}{2}}(-ig_\eta)\, .
\end{equation}

%%   Fig. 11   %%%%%%%%%%%%%%%%%%%%%%%%%%%%%%%%%%%%%%%%%%%%%%%%%%%%%
%%   Fig. 11   %%%%%%%%%%%%%%%%%%%%%%%%%%%%%%%%%%%%%%%%%%%%%%%%%%%%%

%One can see that through a combination of the $t_{44}$ and $\tilde{t}_{44}$
%amplitudes one can determine the $\pi^0(\eta) N^\ast N^\ast$ coupling 
%by
%\begin{equation}
%\label{Cdef1}
%C\vec{\sigma}\cdot\vec{q}\, ' =
%\frac{i\tilde{t}_{44}}{Re (\frac{\partial {t_{44}}}{\partial\sqrt{s}})}\, ,
%\end{equation}
%which is to be calculated around the $N^\ast$ pole. 
%The division by the denominator in eq. (\ref{Cdef1}) 
%has removed the coupling $g^2$ and 
%the square of the $N^\ast$ propagator in eq. (\ref{tm66param}). 

\begin{figure}[tbhp]
\centering
        \footnotesize
\epsfxsize = 15cm
\epsfbox{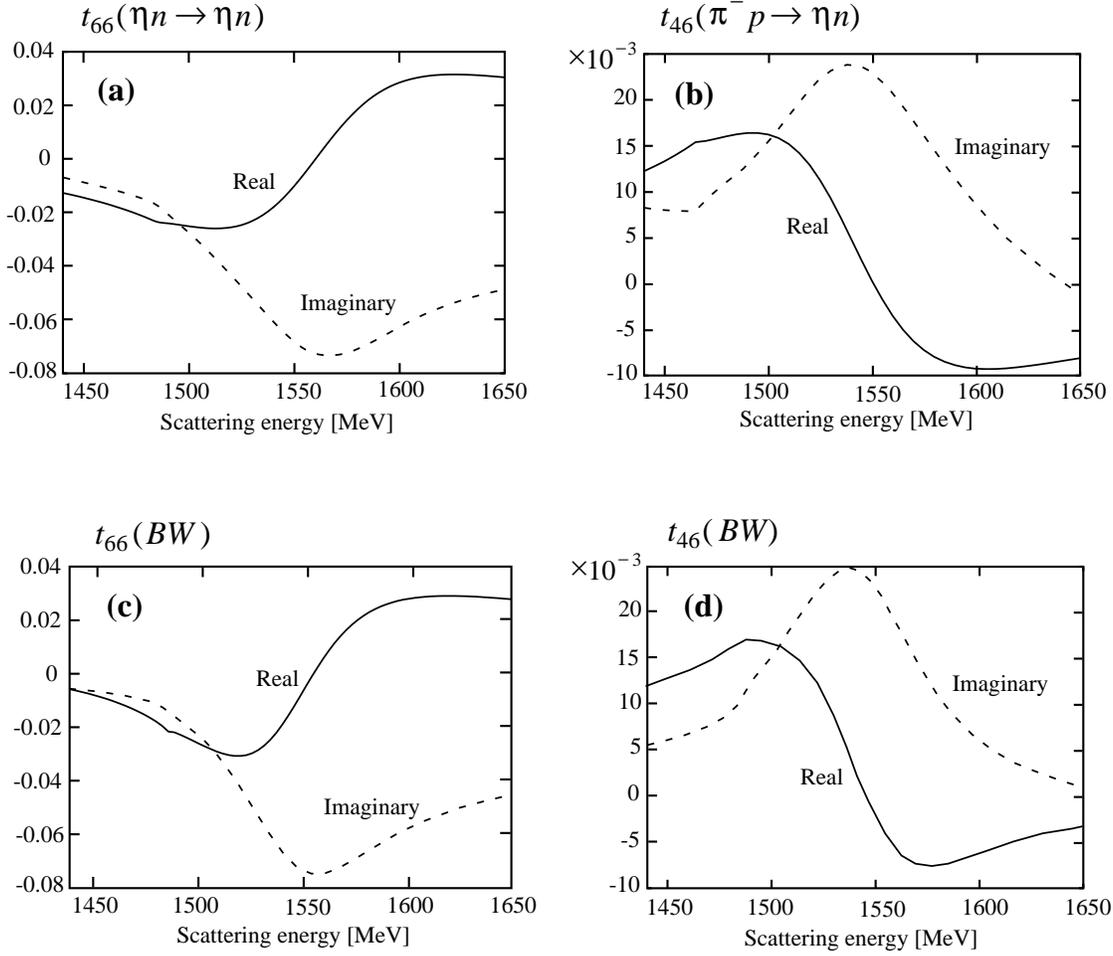}
   \centering 
\caption{Real and imaginary parts of the 
$t_{66}(\eta n \to \eta n)$ (a),  
$t_{46}(\pi^{-} p \to \eta n)$ (b) and the corresponding  
Breit-Wigner 
amplitudes suplemented by the background 
((c) and (d) for eqs. (14) and (15)) 
as functions of the scattering energy $\sqrt{s}$.}
\label{fig:t66t46}
\end{figure}

Our aim is to evaluate the coupling constant $C$ in eq. (\ref{tm66param}). 
This will be accomplished by dividing the amplitude 
by $g_\eta^2$ and the square of the $N^\ast$ propagator.
We should note that the fact
that the $\eta N$ threshold (1486 MeV) is 
below the $N^\ast$ mass (1550 MeV) makes the 
width $\Gamma$ in eqs. (\ref{tm66param}) and 
(\ref{t66param}) strongly energy dependent, particularly
close to the $\eta N$ threshold. 
Hence $t_{66}$ in eq. (\ref{t66param}) differs from a
Breit Wigner distribution close to the $\eta N$ threshold but resembles
it much better at energies above the $N^\ast$ mass. 
This situation is shown in Figs.~\ref{fig:t66t46} (a)-(d), where 
the $t_{66}(\eta n \to \eta n)$ and 
$t_{46}(\pi^{-} p \to \eta n)$
amplitudes of (a) and (b)
are compared with the Breit-Wigner amplitudes of (c) and (d).  

%%   Fig. 12   %%%%%%%%%%%%%%%%%%%%%%%%%%%%%%%%%%%%%%%%%%%%%%%%%%%%%
%%   Fig. 12   %%%%%%%%%%%%%%%%%%%%%%%%%%%%%%%%%%%%%%%%%%%%%%%%%%%%%

%%   Fig. 13   %%%%%%%%%%%%%%%%%%%%%%%%%%%%%%%%%%%%%%%%%%%%%%%%%%%%%
\begin{figure}[tbhp]
\centering
        \footnotesize
\epsfxsize = 15cm
\epsfbox{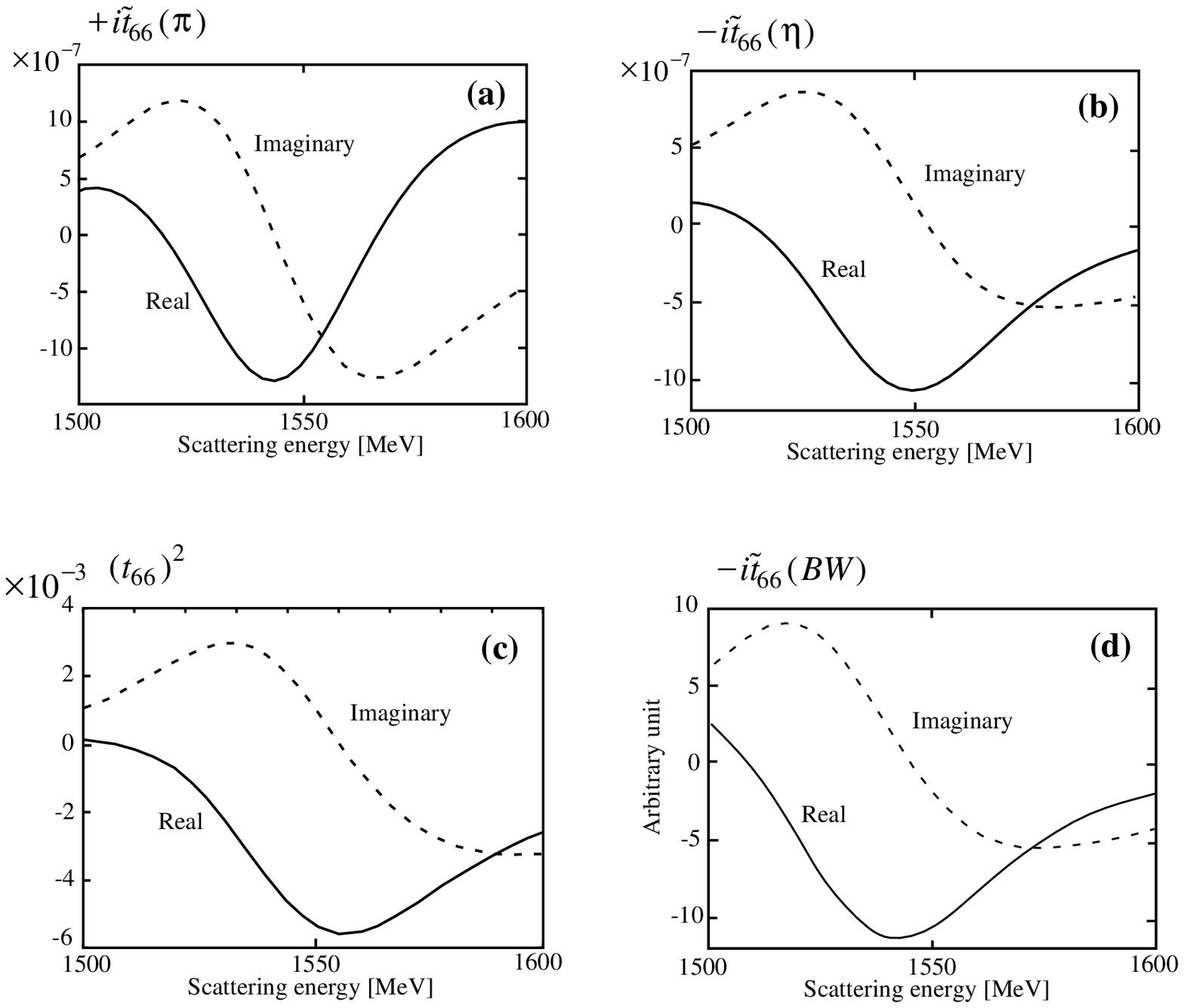}
   \centering 
\caption{Comparison of the amplitude 
$i\tilde t_{66}$ for the pion
coupling (a) and  $-i\tilde t_{66}$ for the eta 
coupling (b), where 
the factor $\vec{\sigma} \cdot \vec{q}\, ^\prime$ has been removed, with 
$t_{66}^2$ (c), as functions
of the scattering energy $\sqrt{s}$. The 
amplitude $-i \tilde t_{66} (BW)$,
obtained assuming a Breit Wigner form for $t_{66}$ (eq. (37)),
is also shown ((d) in arbitrary units).}
\label{fig:tldet66}
\end{figure}
%%   Fig. 13   %%%%%%%%%%%%%%%%%%%%%%%%%%%%%%%%%%%%%%%%%%%%%%%%%%%%%

The amplitude $\tilde{t}_{66}$ both for an external $\pi^0$ and $\eta^0$
behaves approximately  as the square of a Breit Wigner distribution. 
Indeed, 
one should note that all the matrices $t_{6l}$ contain the $N^\ast$ pole. 
This can be seen from eq. (\ref{BStij}) which reads in matrix form
\begin{equation}
t = [1 - VG]^{-1} V
\end{equation}
and all matrix elements have a pole where det$[1 - VG]$ = 0, 
which occurs in the
second Riemann sheet of the complex plane and corresponds 
to the $N^\ast$ pole. 
Hence $\tilde{t}_{66}$ should contain the $N^\ast$ propagator 
squared as implied by eq. (\ref{tm66param}). 
In the vicinity of the complex pole,  
the $N^{*}$ propagator becomes infinite and the $N^{*}$
contribution dominates the process. 
However, if we move on the real axis
for the energy, hence slightly apart from the pole, 
the background in the amplitudes can be nonnegligible, as we found
in eqs. (12-13), and 
the $6\times 6$ amplitudes which can be generated with the 6 states of 
Table~1 have slightly different shapes around 
the $N^\ast$ energy causing the peak of  
$|\Im \tilde t|$ to fluctuate around the $N^\ast$ mass. 
 As a consequence, the
 shapes of $\tilde{t}_{66}$ for an external pion or eta are 
 slightly different and also
 different from $t_{66}^2$, although all of them are qualitatively similar
as can be seen in Fig.~\ref{fig:tldet66}.  
For these reasons, 
%and given the fact that 
the Breit Wigner shape is better
realized to the right of the peak. 
% as we mentioned above, instead of
% taking direct ratios of $\tilde t_{66}$  and 
% $t_{66}^{2}$ at the peak of the resonance
% in order to obtain the constant $C$, 
Therefore, it is better to compare 
$\Re (-i\tilde{t}_{66})$ with $\Re (t^2_{66})$ at the point where they have
their maximum strength 
to the right of the resonance, which happens to be close to the $M^\ast$ pole. 
In this
way we avoid improper comparisons due to the small shifts in the apparent
peak position of the different channels. 
In doing so we minimize the effect of the background
since $\Re (t_{66}^2)$ at its peak comes mostly from ($\Im\, t_{66})^2$. However, considering that the 
background could account for $15\%$ of $\Im\, t$ at its peak, we should accept
uncertainties of the order of $30\%$ in the results from this kind of analysis.
Hence we take
\begin{equation}
 C\vec \sigma \cdot \vec q ^{\; \prime} \simeq 
 g_\eta^2\frac{\Re(-i\tilde{t}_{66})}{\Re (t_{66}^2)}.
\end{equation}

Our technical procedure will
be to evaluate $\tilde{t}_{66}$ from the series of 
Fig.~\ref{fig_rescatt} and apply eq. (40), using the
amplitude ${t}_{66}$ provided by the model of section 2. 
% In this way the $N^\ast$
% propagators generated in the $\tilde{t}_{44}$ amplitude 
% are the same ones generated
% in $t_{44}$ and hence, they and the $\pi^- p N^\ast$ couplings cancel exactly
% in eq. (\ref{tm44param}).
% 
It is not difficult to see that 
the $\tilde{t}_{66}$ amplitude for an external $\pi^0$ is evaluated by 
\be
\label{tm66_t66}
-i\tilde{t}_{66} &=& \sum_{l,m} t_{6l}\, G_l\, D_{lm \pi^0}\, \vec{\sigma}\cdot\
\vec{q}\, ' G_m\, t_{m6}
+ \sum_l t_{6l}\, \tilde{G_l}\, A_l\, \vec{\sigma}\cdot\vec{q}\, 't_{l6} 
\nonumber \\
&+& t_{62} \tilde{G}^\ast A_{\Sigma^0\Lambda\pi^0}\, \vec{\sigma}\cdot
\vec{q}\, ' t_{36} +  t_{63}\tilde{G}^\ast A_{\Sigma^0\Lambda\pi^0}\, 
\vec{\sigma}\cdot\vec{q}\, ' t_{26} \ , 
\ee
where 
\begin{equation}  
        \label{deftildeG}
\tilde{G_{l}}  =  i\int\frac{d^4q}{(2\pi)^4} \frac{M_l}{E_l(\vec{q}\, )}
\frac{M_l}{E_l(\vec{q}\, )}
  \left( \frac{1}{\sqrt{s} - q^0 - E_l(\vec{q}\, ) +i\epsilon}\right)^2 
 \frac{1}{q^2 - m_l^2 + i\epsilon}\, .   
\end{equation}
Here by taking $q'^{\mu}\rightarrow 0$, the square of the baryon
propagator appears. 
The coefficients $D_{lm \pi^0}$ in eq. (\ref{tm66_t66}) are given by 
eq. (\ref{Dllm}), and the coefficients $A_{l}$ are abbreviations for 
$A_{ll\pi^0}$, the diagonal $A_{B'BM}$ coefficients of eq. (22).  
The last two terms in eq. (\ref{tm66_t66}) correspond to Fig.~\ref{fig_rescatt}(c) 
with intermediate $\Sigma^0$, $\Lambda$ or $\Lambda$, $\Sigma^0$
respectively. For the meson baryon propagator $\tilde{G}^\ast$ we take
here the same formula as eq. (\ref{deftildeG}) averaged for the 
$\Lambda$ and
the $\Sigma^0$. Since there is no $\Sigma^0 \Lambda \eta$ coupling, these last
two terms in eq. (\ref{tm66_t66}) 
do not appear for the case of the external $\eta$.

 By taking one of the $M/E$ factors unity, we can
obtain $\tilde{G}_l$ from $G_l$ by differentiating with respect to $\sqrt{s}$
\begin{equation} 
\label{tildeG}
\tilde{G}_l = - \frac{\partial {G_l}}{\partial\sqrt{s}} \ .
\end{equation}

We note that eq.(42) presents a singularity at the meson-baryon threshold, which
disappears if $q^{\prime\,\mu} \neq 0$ because the square of the
baryon propagator is replaced by the product of two different propagators.
This singularity fades away rapidly as one moves away from threshold, as one can 
see in Fig.~\ref{fig:tldet66} where for $\sqrt{s} > 1500$ MeV, only 14 MeV
above the $\eta N$ threshold, there is already no sign of the singular behavior.
It is clear that in this procedure there is some element of arbitrariness
because in order to evaluate $G_l$ we have fixed the cut off arbitrarily
and adjusted the $a_l$ coefficients to the data. Since $a_l$ is a constant it
will not contribute in the derivative of eq. (42). 
In order to estimate the uncertainties from this procedure we conduct a
test consisting in increasing and decreasing the cut off by about 30 \% to
1300 MeV and 700 MeV respectively. 
Accordingly the $a_l$ coefficients of eq.
(10) are changed such that at the energy of the $N^*$ resonance $G_l$ takes the
same value. However, $\tilde G_l$  would now change and so the results 
for $C$.
When we conduct this test we find that the amplitudes in Figs.~\ref{fig:t66t46} 
and 
\ref{fig:tldet66} are barely
changed. The changes are of the order of 5 \% . This gives us confidence 
about the fairness of the method used to evaluate the couplings.

%==================================================================
\section{Results and discussions}
%==================================================================

%---------------------------------
\subsection{Numerical results}
%---------------------------------

As we explained in section 3, we consider the zero charge state for 
$N^\ast(1535)$.  
The evaluation of the $C$ coefficient for $\pi^0 N^\ast N^\ast$ gives us:
\be
%C_{\pi^0 N^\ast N^\ast} = (-8.6 + i 5.6)\cdot 10^{-3} MeV^{-1}\, .
C_{\pi^0 N^\ast N^\ast} = -7.09 \cdot 10^{-4}\; {\rm MeV}^{-1}\;  , 
\label{Cresult}
\ee
with an admitted uncertainty of around 30 $\%$.

The number $C$ in eq. (44) 
should be compared with the equivalent
coupling $\pi^0 N N$, which is given by $-\displaystyle\frac{D + F}{2f}$ 
with $D + F = 
1.26 = g_{A}$.
Hence we find
\begin{equation}
\label{c13}
\frac{C_{\pi^0 N^\ast N^\ast}}{C_{\pi^0 N N}} \sim 0.11\, ,
\end{equation}
which tells us that the $\pi N^\ast N^\ast$ 
coupling of about $10\%$ of the 
$\pi N N$ one and has equal sign. We have checked that the sign is stable
 under changes of the parameters compatible with the data. 
Hence this would rule out the possibility of the mirror assignment. 
Using the ratio of the Goldberger-Treiman relations for $N$ and 
$N^\ast$ 
\be
g_{A}^{*} = \frac{M}{M^{*}} 
\frac{C_{\pi^0 N^\ast N^\ast}}{C_{\pi^0 N N}} g_{A}\, , 
\ee
one can relate the number in eq. (\ref{c13}) to $g_{A}^{*}$:
\be
\label{gA1}
g_{A}^{*} \sim 0.08 \, .  
\ee

As for the $\eta$ coupling we obtain:
\begin{equation}
%  C_{\eta\ N^\ast N^\ast} = (-4.26 - i 3.00)\cdot 10^{-3} MeV^{-1}\, ,
   C_{\eta N^\ast N^\ast} = 5.83\cdot 10^{-4} \; {\rm MeV}^{-1}\ .
\end{equation}
This gives the ratio:
\begin{equation}
  \frac{C_{\eta N^\ast N^\ast}}{C_{\pi^0 N N}} = 0.09\, .
\end{equation}
The strength of the $\eta$ coupling here is less than 
$10\%$ of that of the pion. 
Here, in addition to the $30\%$ uncertainties
due to the non resonant background we should consider 
additional ones as we have discussed above.
  
In order to have some feeling for the strength of the different terms we
summarize here the general trends of the numerics.
\begin{enumerate}
\item  For the case of the pion coupling the contribution of the first 
term in eq. (41), the $GDG$ term, is
about one order of magnitude smaller than the terms involving
$\tilde{G}$. The sum of the two terms in eq. (41) involving the
$\tilde{G^\ast}$ are smaller than the one involving $\tilde{G}$ but of the
same order of magnitude.
 
\item  Given the dominance of the $\tilde{G}$ terms, we find a
cancellation between the terms $\pi^-p$ and $\pi^0 n$ intermediate
states (because of the isospin factor in $\pi N N$, see Table 2)
and hence the contribution is dominated by the $\eta n$ and 
$K^+ \Sigma^-$
intermediate states between which there is still some partial
cancellation.
 
\item  For the case of the eta coupling the contribution of the $\tilde{G}$
terms is also dominant over the $GDG$ combination,  and as we mentioned, the
last two terms in eq. (41) do not appear in this case.
Furthermore, there is no cancellation between the pion intermediate
states (because of the isospin of the $\eta$, see Table 2). 
The dominant terms come from the $K \Sigma$ and $K \Lambda$
intermediate states in which there are some partial cancellations.
\end{enumerate} 

% We can rewrite the coefficients $C_{\pi^0 N^\ast N^\ast}$ and 
% $C_{\eta N^\ast N^\ast}$
% in terms of the $F'$, $D'$ coefficients of SU(3), assuming the
% $N^\ast(1535)$ to be equivalent of the N in an octet of excited baryons.
% By analogy to eq. (\ref{A_bbm}) we would have
% \begin{equation}
%    C_{\pi^0 N^\ast N^\ast} = -\frac{D' + F'}{2f} = -1.3 (\frac{D + F}{2f})
% \end{equation}
% \begin{equation}
%   C_{\eta N^\ast N^\ast} = \frac{1}{\sqrt{3}} \frac{D' + F'}{2f}
%   -\frac{2}{\sqrt{3}}\frac{D' - F'}{2f} = -2.2(\frac{D + F}{2f})
% \end{equation}
% These equations are compatible with
% \begin{equation}
%    D'\simeq 2.4 \, , 
% \end{equation}
% \begin{equation}
%    F'\simeq -0.79 \, , 
% \end{equation}
% from which one could obtain the different $B' B M$ couplings for the excited
% baryon states.
% 
% The value of $D'$ is about three times the value of $D$ for the coupling 
% with the nucleon and the value of $F'$ 
% has a strength
% about twice bigger than $F$ and has opposite sign.
% We should note that the uncertainty in 
% the $N^\ast N^\ast \eta$ coupling  
% has repercussions mostly on $F'$. Indeed, if we decrease

% $C_{N^\ast N^\ast \eta}$ by 30$\%$ $D'$ decreases only by 15$\%$ but $F'$
% is reduced to about half its value in eq. (50).
%---------------------------------
\subsection{Comparison with quark models}
%---------------------------------

It would be interesting to compare the present result with other model 
calculations.  
There are several examples which
appear to support the present result.  

First we discuss the implications of the nonrelativistic (NR) quark 
model, where the negative parity baryons emerge as 
orbitally excited states of $l = 1$ ($p$ state).  
There are two independent states for $1/2^{-}$:
\be
|1\ket &=& [l=1,S=1/2]^{1/2^{-}} \, , \nonumber \\
\label{qst12}
|2\ket &=& [l=1,S=3/2]^{1/2^{-}} \, , 
\ee
where the $p$ state is coupled either by the total spin $S=1/2$ state or 
$S=3/2$.  
In the presence of a tensor force, these two states are
mixed and their linear combinations are interpreted as physical 
resonances:
\be
|N^*(1535)\ket &=& a |1\ket - b |2\ket \, , \\
|N^*(1650)\ket &=& b |1\ket + a |2\ket \, .  
\ee
Here the coefficients $a$ and $b$ are real and satisfy 
$a^{2}+b^{2} = 1$ ($b > 0$).

\begin{figure}[tbhp]
   \centering
        \footnotesize
   \epsfxsize = 7cm
   \epsfbox{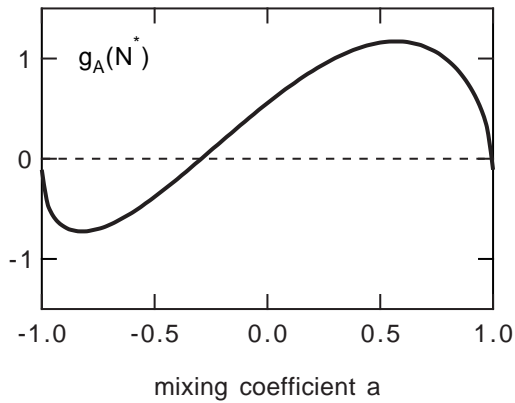} 
%       \centerline{\protect
%       \hbox{
%       \psfig{file=f11ga.eps,height=7.5cm,width=8.cm,angle=0}}}
        \caption{\small $g_A^\ast$ as a function of the mixing coefficient $a$.}
        \label{fig_gAquark}

\end{figure}

The axial vector coupling constant is then defined by 
\be
g^{*}_{A} =  \bra p^{*}(1535), s_{z} =1/2 | 
      \sum_{i=1,2,3} \sigma_{3}^{(i)} \tau_{3}^{(i)} 
          | p^{*}(1535), s_{z}=1/2 \ket \, , 
          \label{gAquark}
\ee
where $\sigma^{(i)}$ and $\tau^{(i)}$ are quark spin and isospin 
matrices for the $i$-th quark and the sum runs over the three quarks.  
The matrix element of eq. (53) is easily computed and the result 
is 
\be
\label{gAvalue}
g^{*}_{A} = \frac{1}{9} 
\left( -a^{2} + 16 a \sqrt{1-a^{2}} + 5 (1-a^{2}) \right) \, .
\ee
In Fig.~\ref{fig_gAquark} we show $g^{*}_{A}$ as a function of the mixing 
coefficient $a$. 
We see that depending on the value of $a$, the coupling constant 
$g^{*}_{A}$ can be either positive or negative.  
In the successful model by Isgur and Karl~\cite{IsgKar}, 
the mixing coefficient takes the value 
$a \sim 0.85$, yielding an axial vector coupling constant 
$g^{*}_{A} \sim 0.87$, which is about an order of magnitude larger
than the present results.  
One should also note, from input of Fig.~\ref{fig_gAquark}, that a
slight  increase of 
the mixing
coefficient above the value 0.85 would reduce drastically $g_A^\ast$.

%%   Fig. 14   %%%%%%%%%%%%%%%%%%%%%%%%%%%%%%%%%%%%%%%%%%%%%%%%%%%%%
%%   Fig. 14   %%%%%%%%%%%%%%%%%%%%%%%%%%%%%%%%%%%%%%%%%%%%%%%%%%%%%

Another interesting case is presented by the large-$N_c$ 
consideration.    
A somewhat nontrivial assumption here is the hedgehog correlation, 
where spin and isospin degrees of freedom
are strongly correlated such that the intrinsic state acquire the 
grand spin $\vec K = \vec J +\vec I = 0$.  
Here $J$ is the total spin, $\vec J = \vec L + \vec S$, and $\vec I$ 
the isospin.  
In the large-$N_c$ limit, quark-quark interactions become small 
as the Hartree approximation becomes exact.  
Hence the hedgehog configuration is assumed for each quark separately.  
For $s$ and $p$ quarks, such hedgehog states are
\be
| h \ket &=& [s,\tau]^K = [1/2,\, 1/2]^{0} \, , \\
| h^* \ket &=&  [[l,s]^j,\tau]^K 
= [ [1,1/2]^{1/2},\, 1/2]^{0} \, .
\ee
Hence the negative parity hedgehog for $N^*(1535)$ 
in the large-$N_c$ limit takes the form
\be
|H\ket = \frac{1}{N_{c}} 
\left(
|h^*\, h \cdots h\ket + \cdots |h \cdots h, h^* \ket
\right) \, .
\ee
Physical baryon states are projected out from the hedgehog state by 
applying the projection operator:
\be
\label{projection}
|B\ket \sim \int d[\omega] \, D^*_B (\omega) \, R^\tau (\omega) \,  
|H\ket \, , 
\ee
where $R^\tau (\omega)$ is the rotation operator in isospin space and 
$D^*_B (\omega)$ the $D$-function for a physical baryon state $B$.  

For $N_c=3$ the integral in eq. (\ref{projection}) yields
\be
|B = 1/2^-\ket = \frac{1}{\sqrt{2}} (|1\ket - |2\ket ) \, ,
\ee
where the states $|1\ket$ and $|2\ket$ are those given
in eqs. (\ref{qst12}).  
Thus the mixing coefficient $a$ in this case is precisely 
$1/\sqrt{2}$ and the corresponding $g_{A}^* = 1.11$.  
We note that this mixing rate is consistent with the result 
obtained in a phenomenological analysis based on the 
large-$N_{c}$~\cite{carone}.

% The last example is provided by the Skyrme model, 
% where the $N^\ast(1535)$ resonance can be 
% constructed dominantly as an $s$-wave eta-nucleon bound 
% state~\cite{pari,hosaka}.  
% When the original Skyrme model is extended minimally to the SU(3) 
% case, the eta which is introduced as a fluctuation field feels 
% an attractive potential around the Skyrme soliton.  
% In the large $N_c$ limit, the soliton solution appears in 
% the leading order and the fluctuation field next to leading order 
% does not affect the shape of the soliton solution.   
% Therefore, the pion coupling which is determined from the tail 
% behavior of the soliton solution takes the same form both for the 

% nucleon and for $N^\ast(1535)$.  
% 
% In the chiral limit $m_\pi = 0$, the strength of 
% the asymptotic soliton profile $F(r) \to b/r^2$ is related to the pion 
% coupling and to the axial coupling through 
% \be
% g_{\pi N^*N^*} &=& \frac{8\pi M^*}{3} f_\pi b \, , 
% \nonumber \\
% g_A^* &=& \frac{8\pi}{3} f_\pi^2 \, b \, , 
% \ee
% which are the same equations as those for the nucleon if the asterisk $*$ 
% is removed~\cite{hostok}.  
% Therefore, using the parameters of Adkins-Nappi-Witten~\cite{ANW}, we 
% find 

% \be
% g_{\pi N^*N^*} = M^*/M_{N} g_{\pi NN} \sim 13.5
% \ee
% and 
% \be
% g_A^* = 0.65 \; (= g_A) \, .
% \ee
% 
% 

{}From these considerations, our present results look very much 
different from the quark model results.  
The discrepancy, however, 
% of the present results with those of quark models is apparent
% but this 
should not be a big surprise since the $N^\ast(1535)$ as well the $
\Lambda(1405)$ resonances are cases known to be somewhat pathological within
standard quark models. Actually we are giving a rather different interpretation
of this resonance as a kind of quasibound meson baryon state in coupled
channels. However, it is also worth noting that other versions of quark models which
account for relativistic corrections in the $\pi q q$ coupling \cite{cano}
lead to values of $C_{\pi^0 N^\ast N^\ast}/C_{\pi^0 N N}$ between 0.06 and 0.1,
depending on the confining potential used, in agreement with our results of eq.
(45).
   
%==================================================================
\section{Conclusions}
%==================================================================

We have evaluated the $N^\ast N^\ast \pi^0$ and $N^\ast N^\ast \eta$
couplings using a chiral unitary approach where the $N^\ast(1535)$ is
generated dynamically. We could generate these couplings in terms  of known
couplings of the $\pi$, $\eta$ to baryons and other mesons as given by the
chiral Lagrangians. The values obtained are about one order of magnitude
smaller than 
the $\pi N N$ coupling. In the case of the $\pi N^\ast N^\ast$ coupling we get
the same sign as the $\pi N N$ one. This is the sign expected for the naive
assignment in the chiral doublet combining mesons and baryons. The
uncertainties in the approach followed here can not reverse this sign, hence
the present calculations appear to rule out the mirror case.

We have compared the present result with other models for $N^*$, 
e.g., the nonrelativistic quark model of Isgur-Karl, 
the large-$N_c$ approach and the relativised quark model of Cano et al.
\cite{cano}.
We have found that all of them predict values for $g_{\pi N^\ast N^\ast}$ 
and $g_A$ with the same sign.  
Hence, $N^*(1535)$ which appears as the first 
resonance state in the pion (and eta) scatterings is likely to 
carry a naive chiral assignment together with the ground state 
nucleon.  However we find absolute values for $g_{\pi N^\ast N^\ast}$ coupling
roughly one order of magnitude smaller than those of quark models, except
the one of Cano et al., which has about the same order of magnitude.

\begin{figure}[tbhp]
   \centering
        \footnotesize
   \epsfxsize = 13cm
   \epsfbox{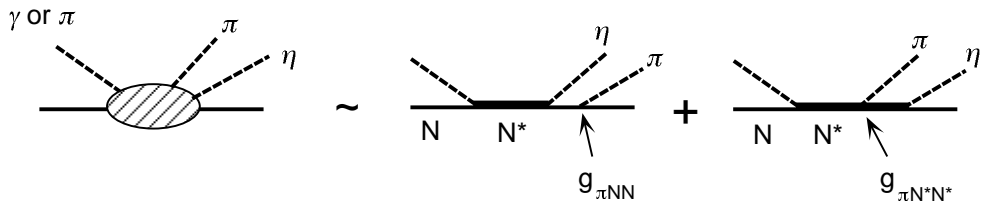} 
%       \centerline{\protect
%       \hbox{
%       \psfig{file=f11ga.eps,height=7.5cm,width=8.cm,angle=0}}}
        \caption{\small A resonance dominant process for the two meson production 
        of $\pi$ and $\eta$. }
        \label{fig_peprod}
\end{figure}

It would be further interesting if we can confirm our theoretical 
implication in experiments.  
Since we can not have a stable target for $N^*(1535)$, the measurement of 
coupling constants will necessarily require somewhat complicated 
processes.  
One such candidate is to measure a two meson production of the pion and 
eta simultaneously~\cite{joh3}.  
The eta meson can be used as a probe for the production of $N^*(1535)$.  
Then, another pion can be produced either at the $N^*(1535)$ or at the 
nucleon with coupling constants $g_{\pi N^*N^*}$ and 
$g_{\pi NN}$, respectively, as illustrated in Fig.~\ref{fig_peprod}.   
Actually, there could be other diagrams which are not shown in 
Fig.~\ref{fig_peprod}.
For instance, the contact term of 
$N^* \pi \eta N$ can be generated by the same method as done here 
by generating the resonance only to the left of the vertex.  
Before performing concrete calculations, however, we 
can expect that from a naive consideration of energy denominators
as emphasized for example by Ochi et al~\cite{takaki}, 
the diagrams shown in Fig.~\ref{fig_peprod} are two 
of the dominant contributions from which we could extract information 
on the sign of the coupling through their interference effect.  
Depending on the sign of the two couplings,  
the two terms can be added either constructively or 
destructively.  
One may use such coherent processes to see the relative sign of the 
two coupling constants~\cite{joh2}. Even if only some bounds on the coupling
could be obtained (in view of the smallness of our predictions) this would 
still be valuable information. 

%%   Fig. 15   %%%%%%%%%%%%%%%%%%%%%%%%%%%%%%%%%%%%%%%%%%%%%%%%%%%%%
%%   Fig. 15   %%%%%%%%%%%%%%%%%%%%%%%%%%%%%%%%%%%%%%%%%%%%%%%%%%%%%

% We have also calculated the corresponding SU(3) coefficients $D'$ 
% and  $F'$, which

% would allow one to calculate the $B' B M$ couplings with the octet of excited
% baryon states to which the $N^\ast(1535)$ belongs and the mesons of the $0^-$
% stable octet. 

% The value obtained for $D'$ is about three times the value of $D$
% while the value of $F'$ is about twice the one of $F$ 
% and of opposite sign.  
% We should, however, note that 
% $F'$ has larger uncertainty than $D'$.  
 
The exercise done here has also served to show how one can technically
implement the coupling of an external pion or eta to a resonance structure.
These techniques can easily be generalized to obtain the coupling of other 
external sources as photons, vector mesons, etc, which would enable one to
evaluate magnetic moments of resonances and other electromagnetic or weak
properties.   
        
%\vspace{3cm}

\subsection*{Acknowledgments}

We are grateful to the COE Professorship program of Monbusho, 
which enabled E. O. to
stay at RCNP to perform the present work.   
Useful discussions and ideas 
from J. A. Oller are much appreciated.  
A.P. would like to acknowledge Prof. R.A. Arndt and N. Kaiser for 
providing her with some experimental data and other helpful 
information.  
A.H. and M.O thank D. Jido for discussions on $\pi$ and $\eta$ 
productions.  
J.C. Nacher would like to acknowledge the hospitality of 
the RCNP of the Osaka University where
this work was done and support from the Ministerio 
de Educaci\'on y Cultura. 
This work is partly supported   
by DGICYT contract number PB96-0753 and PB98-1247, 
and also by DOE Grant DE-FG06-91ER40561.


\begin{thebibliography}{99}   

        \bibitem{coleman}

        S. Coleman, {\it Aspects of Symmetry}, Cambrige University Press, 
        Cambrige (1985).
        
        \bibitem{1} J. Gasser and H. Leutwyler, Nucl. Phys. B250 (1985) 465.
        
        \bibitem{2} U. G. Meissner, Rep. Prog. Phys. 56 (1993) 903.
        
        \bibitem{3} A. Pich, Rep. Prog. Phys. 58 (1995) 563.
        
        \bibitem{4} G. Ecker, Prog. Part. Nucl. Phys. 35 (1995) 1.
        
        \bibitem{5} V. Bernard, N. Kaiser and U. G. Meissner, Int. J. Mod. Phys.
        E4 (1995) 193.
        
        \bibitem{6}  J. A. Oller, E. Oset and J. R. Pel\'aez, Phys. Rev. Lett.
        80(1998)3452; Phys. Rev. D 59 (1999) 71001;
        erratum Phys. Rev. D60 (1999) 099906.  

        \bibitem{7} F. Guerrero and J. A. Oller, Nucl. Phys. B 537 (1999) 459.

        \bibitem{8} J. A. Oller and E. Oset, Nucl. Phys. A620 (1997) 438; 
        erratum Nucl. Phys. A 652 (1999) 407.  
        
        \bibitem{D} J. A. Oller and E. Oset, Phys. Rev. D60 (1999) 074023.  

        \bibitem{Pichmas} G. Ecker, J. Gasser, A. Pich and E. de Rafael, 
        Nucl. Phys. B321 (1989) 311.

        \bibitem{joh1} D. Jido, M. Oka and A. Hosaka,
    Physical Review Letters, 80 (1998) 448.  

        \bibitem{joh2} D. Jido, Y. Nemoto, M. Oka and A. Hosaka,
    TIT Preprint (1998), also hep-ph/9805306, to appear in Nucl. Phys. 
    A.  

        \bibitem{kjo} H. Kim, D. Jido and M. Oka,
    Nucl. Phys. A640 (1998) 77.  

        
        
%       \bibitem{Pichmas} G. Ecker, J. Gasser, A. Pich and E. de Rafael, 
%       Nucl. Phys. B321 (1989) 311

        
        \bibitem{9} N. Kaiser, P. B. Siegel and W. Weise, 
        Phys. Lett. B 362 (1995) 23.

        \bibitem{10} N. Kaiser, P. B. Siegel and W. Weise, 
        Nucl. Phys. A594 (1995) 325.

        \bibitem{11} N. Kaiser, T. Waas and W. Weise, 
        Nucl. Phys. A612 (1997) 297.

        \bibitem{12} E. Oset and A. Ramos, Nucl. Phys. A635 (1998) 99.
%        \bibitem{caro} J. Caro Ramon, N. Kaiser, S. Wetzel and W. Weise,
%        nucl-th/9912035, to be published in Nucl. Phys. A.
%        \bibitem{PDG} Particle Data Group, C. Caso et al, 
%        Eur. Phys. Jour. C3(1998)1

%       \bibitem{D} J. A. Oller and E. Oset, submitted to Phys. Rev. D, 

        
        
        \bibitem{pr} M. Roos, Phys. Lett. B246 (1990) 179.  

        \bibitem{CDD} L. Castillejo, R.H. Dalitz and F.J. Dyson, Phys. Rev. 101 
        (1956) 453.  
        \bibitem{caro} J. Caro Ramon, N. Kaiser, S. Wetzel and W. Weise,
        nucl-th/9912035, to be published in Nucl. Phys. A.
	
	\bibitem{PDG} Particle Data Group, C. Caso et al, 
        Eur. Phys. Jour. C3(1998)1
        


%       \bibitem{15} R. A. Arndt, A. M. Green, R. L. Workman and S. Wycech, 
%       Phys. Rev. C58 (1998) 3636.
        \bibitem{chiang} H. C. Chiang, E. Oset and L. C. Liu, Phys. Rev. C44
        (1991) 738
        

        \bibitem{meiss} U. G. Meissner, E. Oset and A. Pich, Phys. Lett. B353 
        (1995)  161.

        \bibitem{wein} S. Weinberg, Phys. Rev. Lett. 18 (1967) 188.

        \bibitem{IsgKar}
     N. Isgur and G. Karl, Phys. Lett.  72B (1977) 109; {\it ibid.} 
     74B (1978) 353; Phys. Rev.  D18 (1978) 4187; {\it ibid.} D20 (1979) 1191 

        

        \bibitem{carone} C.D.~Carone, H. Georgi, L. Kaplan and D. Morin, 
           Phys.Rev. D50 (1994) 5793.

        
%        \bibitem{pari}  G. Pari, Phys. Lett. B261 (1991) 347.
        
%        \bibitem{hosaka} A. Hosaka, Phys. Lett. B293 (1992) 23; 
%                        {\it ibid.} B306 (1993) 207.  
                        
%        \bibitem{hostok}  A. Hosaka and H. Toki, 
%        Phys. Reports, 277 (1996) 65. 

%    \bibitem{ANW} 
%    G.S. Adkins, C.R. Nappi and E. Witten, Nucl. Phys. B228 (1984) 552. 

  	\bibitem{cano} F. Cano, P. Gonz\'alez, B. Desplanques, S. Noguera, 
	Z. Phys. A359 (1997) 315-319; F. Cano, P. Gonz\'alez, S. Noguera, 
	B. Desplanques, Nucl. Phys. A603 (1996) 257-280; F. Cano,  
	private communication
  
  
        \bibitem{joh3} D. Jido, M. Oka and A. Hosaka, in preparation.    

        \bibitem{takaki}  K. Ochi, M. Hirata and T. Takaki, 
    Phys. Rev. C56 (1997) 1472, and private communication from T. Takaki.  
        
\end{thebibliography}
\end{document}